\documentclass[12pt]{iopart}
\usepackage{graphicx}
\usepackage{natbib}
\usepackage{cite}
\usepackage[utf8]{inputenc}
\usepackage[T1]{fontenc}
\usepackage{url}
\usepackage{hyperref}
\usepackage{bm}
\usepackage{tabularx}
\usepackage{xcolor}
\usepackage{iopams}

\usepackage{fancyhdr}
\expandafter\let\csname equation*\endcsname\relax

\expandafter\let\csname endequation*\endcsname\relax

\usepackage{amsmath}

\begin{document}

\title[]{Generation and annihilation of three dimensional magnetic nulls in extrapolated solar coronal magnetic field: Data-based Implicit Large Eddy Simulation}

\author{Yogesh Kumar Maurya$^{1,2}$, Ramit Bhattacharyya$^1$, David I. Pontin$^3$}

\bigskip

\address{$^1$Udaipur Solar Observatory, Physical Research Laboratory, Udaipur, Rajasthan 313001, India.}
\address{$^2$ Department of Physics, Indian Institute of Technology, Gandhinagar, Gujarat 382055, India.}
\address{$^3$ School of Information and Physical Sciences, University of Newcastle, Callaghan, NSW 2308, Australia.}

\ead{yogeshk@prl.res.in}
\vspace{10pt}
\begin{indented}
\item[] \today
\end{indented}

\begin{abstract}
Three-dimensional (3D) magnetic nulls are the points where magnetic field vanishes and are preferential sites for magnetic reconnection: a process which converts magnetic energy into heat and accelerates charged particles along with a rearrangement of magnetic field lines. In the solar corona, the reconnections manifest as coronal transients including solar flares, coronal mass ejections and coronal jets. The nulls are generally found to be collocated with complex active regions on the solar photosphere. Extrapolation of magnetic field from corresponding photospheric magnetogram indicate an abundance of these nulls in the solar atmosphere. Nevertheless, their generation is still not well understood. Recently, \citet{2023PhPl...30b2901M} have demonstrated magnetic reconnection to be a cause for generation and annihilation of magnetic nulls through magnetohydrodynamics simulation, where the initial magnetic field is idealized to have a single proper radial null. This article further extends the study in a more realistic scenario where the initial magnetic field is constructed by extrapolating photospheric magnetogram data and hence, incorporates field line complexities inherent to a complex active region. For the purpose, the active region NOAA $11977$ hosting a C6.6 class flare is selected. The simulation is initiated using non-force-free extrapolated magnetic field from the photospheric vector magnetogram at around $02:48:00$ UT on 17 February 2014, 16 minutes before the flare peak. The generation, annihilation and dynamics of nulls are explored by a complimentary usage of trilinear null detection technique and tracing of magnetic field line dynamics. It is found that the nulls can spontaneously generate/annihilate in pairs while preserving the topological degree and can have observational implications like footpoint brightenings. Magnetic reconnection is found to be the cause of such generation and annihilation.
\end{abstract}

\maketitle

\section{Introduction}
Three-dimensional (3D) nulls are locations in space where the magnetic field is zero. Contemporary researches in theory, observation and numerical simulation have shown the importance of 3D magnetic nulls in natural phenomena involving magnetic reconnection: a process where magnetic field line connectivity changes along with generation of heat and acceleration of charged particles. 3D nulls are preferential sites for magnetic reconnections \citep{priest_forbes_2000, birn_priest_2007,pontin2007, 2009ARA&A..47..291Z, 2010RvMP...82..603Y, 2022LRSP...19....1P} which, in turn, are responsible for various transient events in the solar corona---including persistent null point brightening \citep{2023NatCo..14.2107C}, solar flares \citep{2000ApJ...540.1126A,  2001ApJ...554..451F, 2009ApJ...700..559M}, coronal jets \citep{wyper2018,2019ApJ...875...10N} and coronal mass ejections \citep{Lynch_2008, 1996SoPh..169...91L}. 

The theory of 3D nulls is well-established and the following properties are worth mentioning. Straightforward manipulation of the ideal induction equation for an incompressible fluid shows
\begin{equation}
    \label{dbdt}
    \frac{d\textbf{B}}{dt}=0,
\end{equation}
at the null point \citep{doi:10.1063/1.871778}; the $d/dt$ being the Lagrangian derivative. Consequently a 3D null maintains identity during evolution, enabling its tracing in numerical simulations. Moreover, the net topological degree  of a system consisting of a number, $\textnormal{N}_{0}$, of nulls is defined by \citet{1992JCoPh..98..194G} and \citet{2005LRSP....2....7L} as
\begin{equation}
\label{td}
 \textnormal{D} = \sum_{\textnormal{N}_{0}}\textnormal{Sign (det} (\nabla \textnormal{B}|_{x_{\textnormal{N}_{0}}} )),   
\end{equation}
and remains conserved \citep{doi:10.1063/1.871778, 2022LRSP...19....1P} in any 
evolution. Any credible simulation targeted to explore generation or 
annihilation of nulls must satisfy this stringent conservation. Additionally, nulls that enter or exit the computational domain needs to be accounted for as they may seemingly violate the conservation of topological degree.

Magnetic topology of 3D nulls can be put in perspective through a 3D generalization of a two-dimensional (2D) separatrix: the line in 2D that segregates magnetic field lines having separate connectivities. As could be expected, in 3D, the sepratrix lines get replaced by surfaces -- called separatrix surfaces, or just separatrices -- separating sub-volumes having disjoint field line connectivities. To maintain different subvolumes connection-wise disjoint, the separatrices need to be magnetic flux surfaces with only tangential field lines.  Consequently, If two such separatrices intersect, they intersect along a line having two magnetic nulls at the end points. This line is called a separator. The concept can be straightforwardly applied to magnetic structures in the solar corona. Traditionally, such field line topologies are often realized in the solar corona when a parasitic polarity region emerges inside a larger opposite polarity region on the photosphere. In such a case, topological structure of a 3D null point defines a dome-like separatrix, the fan, and two singular field lines, the spines, originating from the null point. With the onset of reconnection at the null, magnetic field lines are transferred across the separatrices from one magnetic domain to another \citep{2013ApJ...774..154P}.

Recent studies \citep{Aulanier_2000, Zhao_2008, 2009SoPh..254...51L, 2002SoPh..207..223S, 2003ApJ...592..597M} have found these nulls to be abundant in the solar atmosphere. Moreover, the majority of these studies consider only equilibrium fields, while more realistic, dynamically evolving fields typically contain many more nulls \citep[e.g.][]{2010PhPl...17i2903H}. However, their origin  is yet to be systematically explored.  Toward such exploration, an Implicit Large Eddy Simulation (ILES) carried out by \citet{2023PhPl...30b2901M} has indicated magnetic reconnection to be a plausible cause for generation and annihilation of magnetic null pairs. In the simulation, the initial magnetic field had a current-free radial null, the spine of which was deformed by a prescribed sinusoidal flow, to achieve primary reconnection.  Further evolution shows spontaneous generation and annihilation of sparsely located null pairs through further reconnections. In reality, solar active regions are characterized by more complex magnetic topology and hence, the corresponding 3D nulls are expected to follow the same complexity. Additionally, a tailor-made flow to initiate coronal field reconnection as considered by \citet[]{2023PhPl...30b2901M} cannot be expected. It is then imperative to study the feasibility of spontaneous null generation/annihilation in a realistic system like the solar corona where magnetic reconnection in all scales is ubiquitous, and if found, to identify associated brightening. Such studies can have direct relevance to chromospheric and coronal heating. To achieve this objective, in this paper a novel approach is adopted by carrying out a data-based ILES for a flaring active region where the initial coronal magnetic field is computed by extrapolating photospheric vector magnetogram data.

The organization of the paper is as follows: Section \ref{sec:extrapolation} describes the extrapolation technique along with the rationale behind selecting the particular active region and section \ref{sec:G_Equation} summarizes the governing magnetohydrodynamic (MHD) equations while briefly describing salient features of the numerical model. Sections \ref{sec:numerical_setup} and \ref{sec:MHD simulations} are dedicated to numerical setup and simulation results respectively. Section \ref{sec:summary} summarizes the findings.

\section{Non Force Free Field Extrapolation of NOAA AR 11977} 
\label{sec:extrapolation}
\begin{figure}[ht]
\includegraphics[width=1\linewidth]{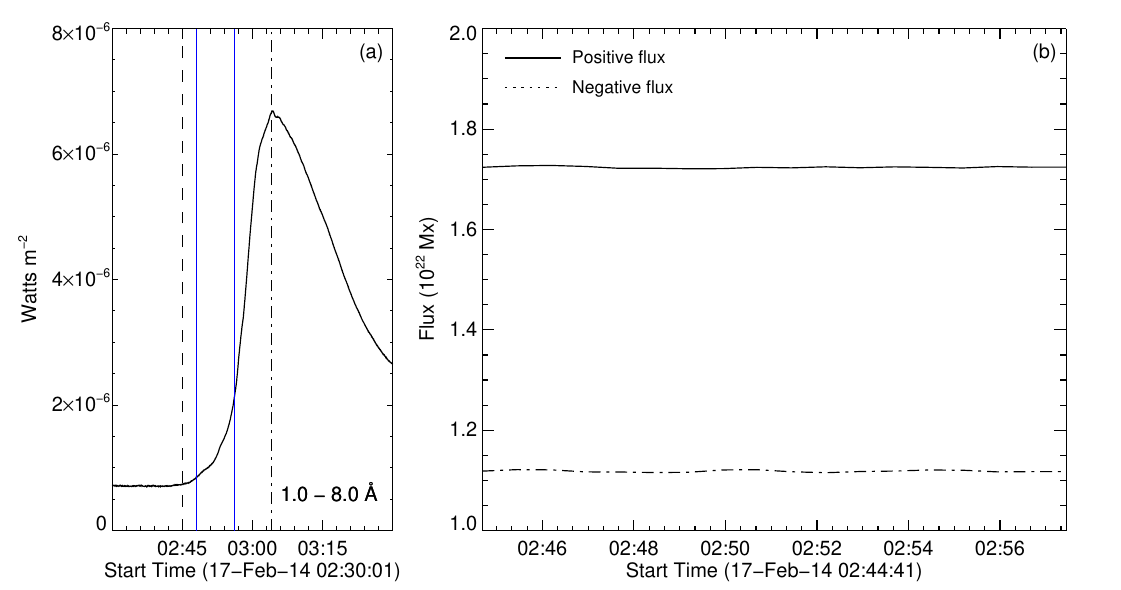}
\caption{Panel (a) depicts the Geostationary Operational Environmental Satellite (GOES) soft X-ray flux over the duration of the flare in the $1-8$ \AA channel. This graph illustrates a gradual increase in intensity starting around $02:45$ UT (marked by the dashed vertical line), with the peak occurring at $03:04$ UT (dash-dot vertical line). Our simulations covers the time range from 02:48 to 02:56 UT as marked by two blue vertical solid lines during rising phase of flare. The photospheric flux variation for approximately 13 minutes, starting from $02:44:41$ UT, is shown in panel (b), where the solid line represents positive flux and the dashed line represents negative flux.}
\label{magnetic_flux_variation}
\end{figure}

The C$6.6$ class eruptive flare that occurred on February $17, 2014$, emanating from active region NOAA AR$11977$ at heliographic coordinates $S13W05$, has been chosen for analyses for three key reasons: (a) its proximity to the solar disk center, which ensures minimal errors in the observed photospheric magnetic field, (b) the photospheric magnetic flux integrated across the active region remains approximately constant throughout the flare, allowing for the application of a line-tied boundary condition to simplify simulations, (c) contemporary observations in multiwavelengths \citep{mitra, Ibrahim_2022}. In Figure \ref{magnetic_flux_variation}(a), the Geostationary Operational Environmental Satellite (GOES) soft X-ray flux in the $1-8$ \AA  ~channel is depicted over the duration of the flare. The plot illustrates a gradual increase in intensity starting around $02:45$ UT (marked by dashed vertical line), with the peak occurring at $03:04$ UT (dash-dot vertical line).  Magnetic field line dynamics in the time interval from 02:48 to 02:56 UT, as indicated by the two solid blue vertical lines during the flare's ascending phase, is numerically explored here. Notably, panel (b) depicts the evolution of the horizontally averaged positive (solid line) and negative (dashed line) photospheric magnetic flux obtained from the $hmi.M45$ series of the Helioseismic Magnetic Imager (SDO/HMI), as described by \citet{2012SoPh..275..229S} and \citet{2012SoPh..275..207S}. The flux plot covers approximately $13$ minutes, commencing around $02:44:41$ UT. The magnetic flux remains relatively stable during the flare, with both positive and negative fluxes exhibiting relative changes well within $1$ \%. 

The MHD simulation carried out here uses extrapolated magnetic field from a vector magnetogram as an initial magnetic field. For extrapolation, the Active Region AR$11977$ at $02:48:00$ UT on February $17, 2014$ is selected based on its prior analysis in terms of identifying the primary reconnection site \citep{2022SoPh..297...91A}. The corresponding magnetogram is obtained from the Helioseismic Magnetic Imager (HMI; \citet{2012SoPh..275..229S}) on board the Solar Dynamic Observatory (SDO) and is taken from the `hmi.sharpcea720s' data series. This data series provides the magnetic field on a Cartesian grid, which is initially remapped onto a Lambert cylindrical equal-area (CEA) projection and then transformed into heliographic coordinates \citep{2014SoPh..289.3549B}. The extrapolation utilizes a non-force-free field (NFFF) extrapolation model for the magnetic field, ${\bf B}$, obtained by minimization of total energy dissipation rate, described in \citet{2007SoPh..240...63B}. The NFFF $\textbf{B}$ obeys a double-curl Beltrami equation for which a solution can be attempted by  expressing it as
\begin{equation}
    \textbf{B} = \textbf{B}_{1} + \textbf{B}_{2} + \textbf{B}_{3};\qquad \nabla \times \textbf{B}_{i} = \alpha_{i} \textbf{B}_{i},
\end{equation}
where $i = 1, 2, 3$ \citep{2008SoPh..247...87H, 2008ApJ...679..848H}. Here, each sub-field $\textbf{B}_{i}$ represents a linear force-free field (LFFF) characterized by specific constants $\alpha_{i}$. Without a loss of generality, a selection  $\alpha_{1} \neq \alpha_{3}$ and $\alpha_{2} = 0$ can be made, rendering $\textbf{B}_{2}$ a potential field. Subsequently, an iterative approach is employed to determine the optimal pair $\alpha = {\alpha_{1}, \alpha_{3}}$, which finds the pair by minimizing the average deviation between the observed ($\textbf{B}_{t}$) and the calculated ($\textbf{b}_{t}$) transverse field on the photospheric boundary. Effectively, the metric 
\begin{equation}
    \textnormal{E}_{n} = \bigg(\sum_{i=1}^{M}|\textbf{B}_{t,i} - \textbf{b}_{t,i}|\times |\textbf{B}_{t,i}| \bigg)\bigg / \bigg(\sum_{i=1}^{M}|\textbf{B}_{t,i}|^{2} \bigg),
\end{equation}
where $M = N^{2}$ represents the total number of grid points on the transverse plane is iteratively minimized \citep{2018ApJ...860...96P}. To achieve an optimal value of $E_{n}$, a corrector potential field to $\textbf{B}_{2}$ is further derived from the difference transverse field, i.e., $\textbf{B}_{t} - \textbf{b}_{t}$, and added to the previous $\textbf{B}_{2}$ in anticipation of an improved match between the transverse fields, as measured by the $E_{n}$. The grid points are weighted with respect to the strength of the observed transverse field to minimize the contribution from the weaker fields (see \citet{2008SoPh..247...87H}; \citet{2010JASTP..72..219H}, for further details).

\begin{figure}[ht]
\includegraphics[width=0.90\linewidth]{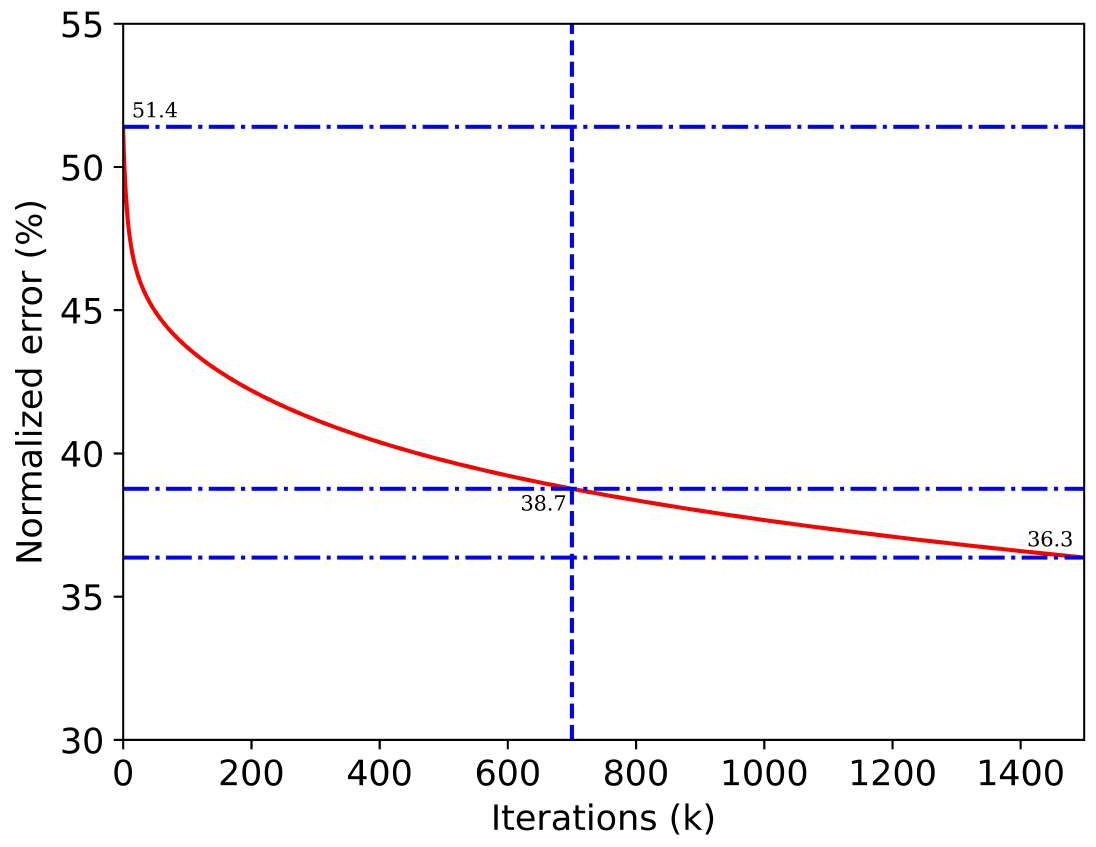}
\caption{The plot illustrates the variation in minimized deviation ($\textnormal{E}_{n}$) with the number of iterations (k) for Non-Force Free Field extrapolation. This deviation decreases monotonically and saturates approximately at $\approx 36.3 \%$ for $1500$ iterations.}
\label{Enplot}
\end{figure}

\begin{figure}[!b]
\includegraphics[width=1.0\textwidth]{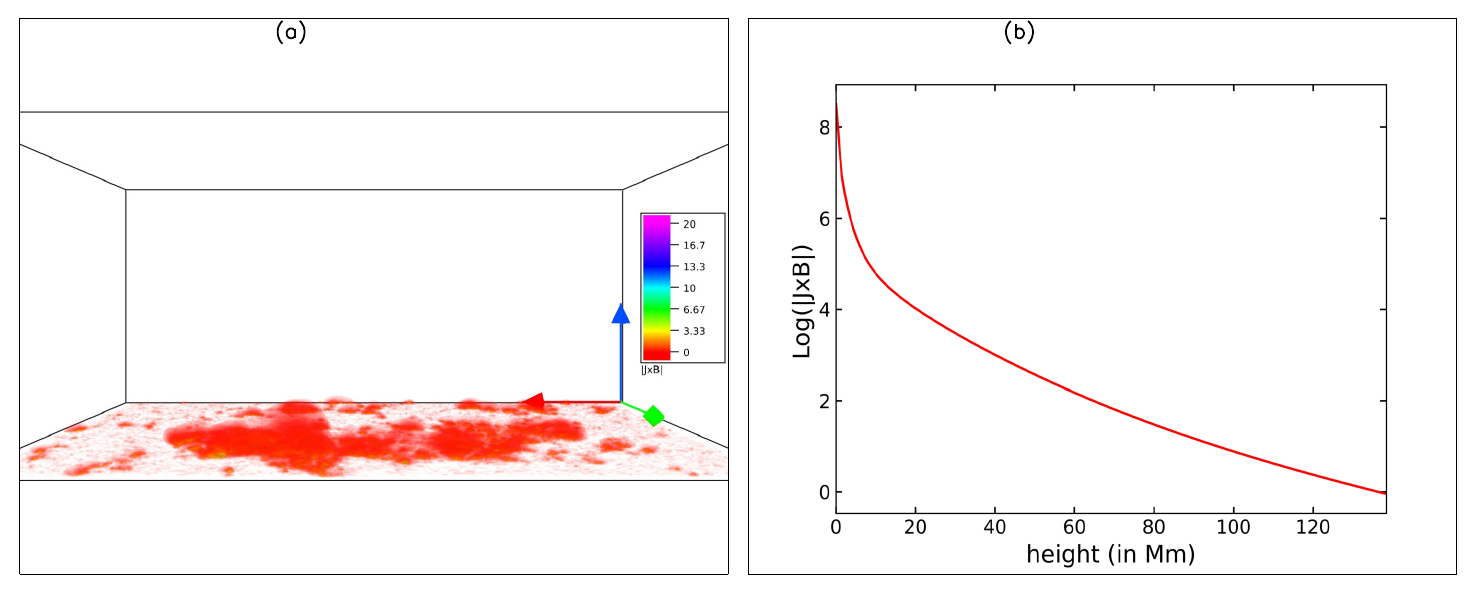}
\caption{Panel (a) of the Figure depicts the direct volume rendering (DVR) of the magnitude of Lorentz force, showing the presence of the Lorentz force at lower heights. To further corroborate this observation, the logarithmic variation in the horizontally averaged strength of Lorentz force with height ($z$) is shown in panel (b). As expected, the logarithmic value of horizontally averaged Lorentz force decreases with height. Notably, the Lorentz force density is nonzero near the photosphere and nearly vanishes at coronal heights, similar to the typical description of the solar corona.}
\label{Lorentz_force}
\end{figure}

To optimize computational cost while preserving the original magnetic morphology, the magnetogram having dimension $896 \times 512$ pixels is re-scaled to the dimension $448 \times 256$ pixels in $x$ and $y$ -directions, respectively. The NFFF extrapolation is carried out on the re-scaled computational grid which corresponds to physical dimensions of $\sim 324.8\textnormal{Mm} \times 185.6 \textnormal{Mm} \times 139.2 \textnormal{Mm}$ in the x, y, and z directions, respectively, and variation of $\textnormal{E}_{n}$ with iteration number is shown in the Figure \ref{Enplot}. As expected, the overall field line morphology is found to be identical to the one depicted in \citep{2022SoPh..297...91A}. Panel (a) of Figure \ref{Lorentz_force} illustrates direct volume rendering (DVR) of the Lorentz force amplitude, showing it to be dominant at lower heights. To further corroborate, the logarithmic variation of horizontally averaged Lorentz force with pixel height $z$ is shown in the panel (b) of Figure \ref{Lorentz_force}, showing a decreasing trend. Effectively the NFFF model treats the corona as reasonably force-free, with non-zeor magnetic forces being localized at the photosphere/lower heights, in agreement with the general expectation of a force-free corona \citep{2020ApJ...899...34L, 2020ApJ...899L...4Y}.

\section{Governing Equations and Numerical Model} \label{sec:G_Equation}
The simulations are carried out using the magnetohydrodynamic numerical model EULAG-MHD  \citep{SMOLARKIEWICZ2013608} assuming the plasma to be thermodynamically inactive, incompressible, and  perfectly electrically conducting. The dimensionless governing equations are
\begin{eqnarray}
\label{ns1}
 &&  \frac{\partial \textbf{v}}{\partial t} + (\textbf{v} \cdot \nabla) \textbf{v} 
= -\nabla p + (\nabla \times \textbf{B})\times \textbf{B} + \frac{1}{R^{A}_{F}} \nabla^{2} \textbf{v} ,\\  
&& \nabla \cdot \textbf{v} = 0,\\
&& \frac{\partial \textbf{B}}{\partial t} = \nabla \times (\textbf{v}\times \textbf{B}), \\
&& \nabla \cdot \textbf{B}  = 0,
\end{eqnarray}

\noindent achieved with 
\begin{eqnarray}
    \label{normalization}
    && \textbf{B}\rightarrow \frac{\textbf{B}}{B_{0}}, \textbf{v}\rightarrow \frac{\textbf{v}}{V_{A}}, L \rightarrow \frac{L}{L_{0}}, t \rightarrow \frac{t}{\tau_{a}}, p \rightarrow \frac{p}{\rho_{0}V^{2}_{A}},
\end{eqnarray}
where $R^{A}_{F} = \frac{V_{A} L}{\nu}$ is an effective fluid Reynolds number with $V_{A}$ as the Alfv\'en speed and $\nu$ as the kinematic viscosity.
The $B_{0}$ and $L_{0}$ are characteristic values of the system under consideration whereas $\rho_{0}$ represents the constant mass density.
Although not strictly applicable in the solar corona, the incompressibility is invoked in other works also \citep{1991ApJ...383..420D, 2005A&A...444..961A}. 
With details in \citet{SMOLARKIEWICZ2013608} (and references therein), essential features of the EULAG-MHD model are highlighted here. Central to the model is the spatiotemporally second-order accurate, non-oscillatory, forward-in-time, multidimensional, positive-definite advection transport algorithm MPDATA \citep{https://doi.org/10.1002/fld.1071}. Two properties of MPDATA are of importance for the simulation presented here. First, the governing prognostic Equations (5) and (7) are both solved in the Newtonian form with total derivatives of dependent variables and the associated forcings forming the left- and right-hand side, respectively; see Section 2.1 in \citet{SMOLARKIEWICZ2013608} for a discussion. This guarantees identity of null preservation as the associated forcing of the induction equation vanishes at the nulls to the accuracy of the field solenoidality (8), which is high \citep{SMOLARKIEWICZ2013608}. 
The second important property is the proven dissipative nature of the MPDATA \citep{2003PhFl...15.3890D,2011ApJ...735...46R, 2016AdSpR..58.1538S}. This dissipation is intermittent and adaptive to the generation of under-resolved scales in field variables for a fixed grid resolution. Using this dissipation property, the MPDATA removes under-resolved scales by producing locally effective residual dissipation of the second order in grid increments, enough to sustain the monotonic nature of the solution in advective transport. The consequent magnetic reconnection is then in the spirit of ILESs that mimics the action of explicit subgrid-scale turbulence models, whenever the concerned advective field is under-resolved, as described in \citet{2006JTurb...7...15M}. Such ILESs performed with the model have successfully simulated regular solar cycles by \citet{2010ApJ...715L.133G} and \citet{2011ApJ...735...46R}, with the rotational torsional oscillations subsequently characterized and analyzed in \citet{2013SoPh..282..335B}. The simulations carried out here also utilize the ILES property to initiate magnetic reconnections already shown by \citet{2016ApJ...830...80K}.

\section{Numerical Setup} \label{sec:numerical_setup}
The active-region cutout is mapped on a grid having 448 $\times$ 256 $\times$ 192 pixels resolved on a computational grid of x $\in \{-0.875, 0.875\}$, y $\in \{-0.5, 0.5\}$, and z $\in \{-0.375, 0.375\}$ in a Cartesian coordinate system, spanning a physical domain of $324.8 \textnormal{Mm} \times 185.6 \textnormal{Mm} \times 139.2 \textnormal{Mm} $ in $x$, $y$ and $z$ directions. The dimensionless spatial step sizes are $\Delta x$ = $\Delta y$ = $\Delta z$ $\approx$ 0.0039 ($\approx 725$ km) and the dimensionless time step is $\Delta t$ = $2 \times {10}^{-3}$ ($\approx 1.936$ s). The initial state is motionless ($\textbf{v} = 0 $) and the initial magnetic field is provided from the NFFF extrapolation. The non-zero Lorentz force associated with the extrapolated field pushes the plasma to generate dynamics. The mass density is set to $\rho_{0}=1$. The effective fluid Reynolds number is set to 5000, which is 5 times smaller than the coronal value of $\approx $ 25000 (calculated using kinematic viscosity $\nu  = 4 \times {10}^{9}$ \text{m$^{2}$ s$^{-1}$} in solar corona, p. 791 of \citet{2005psci.book.....A}). Without any loss in generality, the reduced \text{R$_{F}^{A}$} can be realized as a reduction in computed Alfv\'en speed, \text{V$_{A}|_{\text{computed}}$} $\approx$ 0.14 $\times$ \text{V$_{A}|_{\text{corona}}$}. The Alfv\'en speeds are estimated with characteristic scales L$_\textnormal{computational} = 139.2 \textnormal{Mm}$ for the computational domain and L$_\textnormal{corona} = 100\textnormal{Mm}$ for typical corona. The results presented herein pertain to a run for 250$\Delta t$, which, with $\tau_{\text{A}} \approx 9.68 \times 10^{2}$s, correlates to an observation time of $\approx$ 8 minutes. The interval is marked by the two vertical blue lines of goes X-ray flux curve during ascending phase of the flare (Figure \ref{magnetic_flux_variation}). The reduced \text{R$_{F}^{A}$} slows down the dynamics and does not affect the reconnection mechanism or its consequence, reducing the computational costs and making data-based simulations computationally less costly---realized by \citet{2016NatCo...711522J}. Nevertheless, such reduction in Alfv\'en speed will directly affect the wave dynamics, which is overlooked in this paper in favor of the reconnection dynamics.


\section{Simulations results} \label{sec:MHD simulations}
\begin{figure}[!b]
\includegraphics[width=1.0\textwidth]{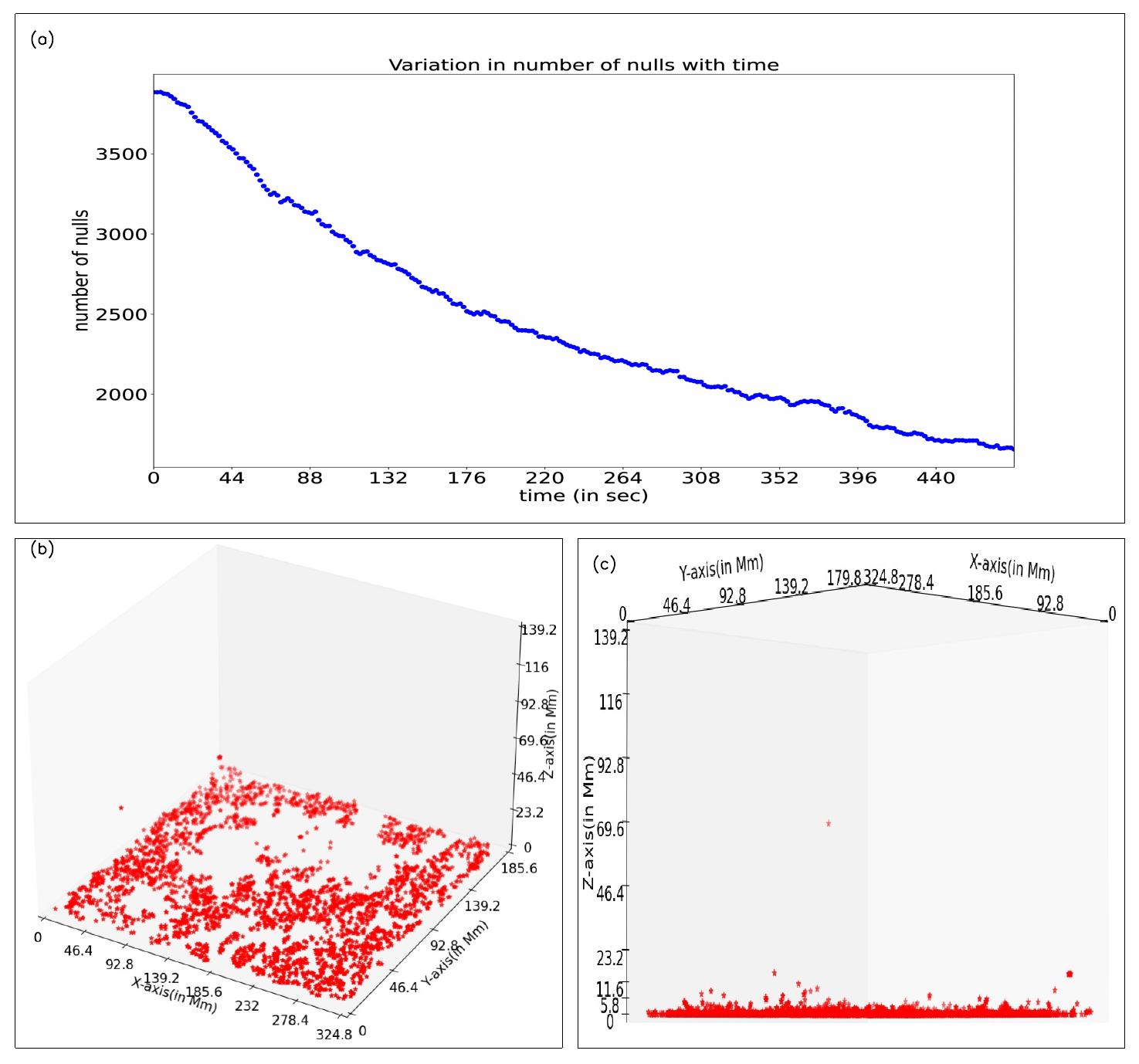}
\caption{The panel (a) illustrates the overall evolution of the number of nulls over time. The vertical axis represents the number of nulls, while the horizontal axis represents time (in seconds). As time progresses, the overall number of nulls decreases. At $t = 0$s, there are approximately $4000$ nulls are present, which are distributed as depicted in panels (b) and (c) (panel (c) is from a different angle to show the distribution). The size of the box is $324.8 \textnormal{Mm}$, $185.6 \textnormal{Mm}$ and $139.2 \textnormal{Mm}$ in x-, y- and z-direction respectively. The nulls are primarily located in the lower solar atmosphere, with some also found in the higher solar atmosphere (refer to panel (c)).}
\label{num_null_variation}
\end{figure}

A modified trilinear null detection technique is used to detect magnetic nulls which, now additionally provides the topological degree (TD) of the nulls as either $+1$ or $-1$ based on the sign of the determinant of $\nabla B|_{\textnormal{null}}$. In the simulation, the overall number of nulls is found to decrease with time (see panel (a) of Figure \ref{num_null_variation}). At the beginning ($t = 0$~s), around 4000 nulls are found to be primarily located in the lower solar atmosphere, as depicted in panels (b) and (c) of Figure \ref{num_null_variation}, with some also found in the higher solar atmosphere. Focus is set on generation/annihilation of three types of null pairs, selected a posteriori; based on their traceability, diversity and tractability of corresponding field lines that reconnect. Throughout the simulation, many such generation and annihilation process of each type occur, as the total number of nulls varies as shown in Figure \ref{num_null_variation}(a). In the remainder of this section we describe one characteristic example of each of the three types. The three pairs of nulls are positioned away from the computational boundary and as per their topology and dynamics are labeled as (i) radial-radial-pair, (ii) radial-spiral-pair-1 and, (iii) radial-spiral-pair-2; listed sequentially as per their generations. The dynamics of field lines leading to generation/annihilation of each pair is discussed below.

\subsection{Radial-radial-pair} \label{subsec:radial}
\begin{figure}[!b]
\includegraphics[width=0.9\textwidth]{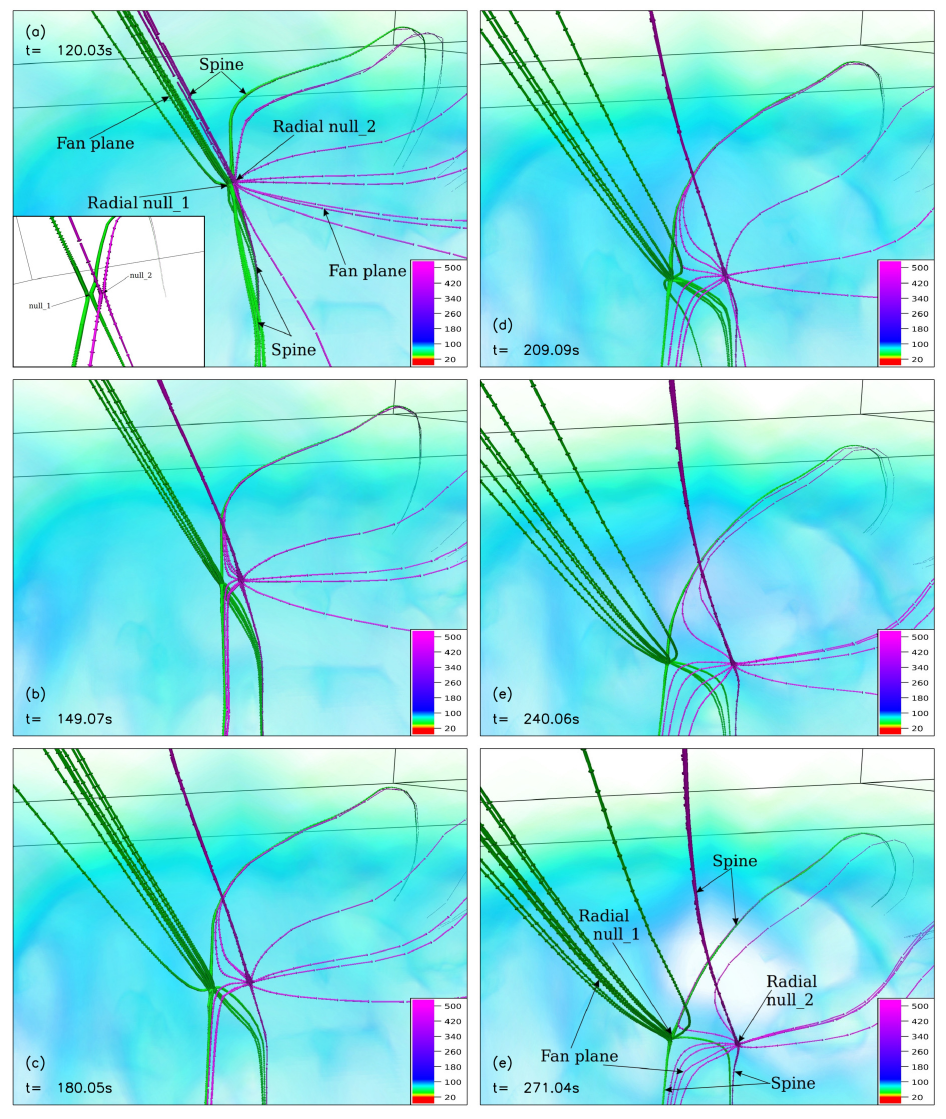}
\caption{The figure illustrates the evolution of radial nulls over time. Nulls are traced in time, and field lines are drawn at their locations. The variation in magnitude of current intensity (identified by the Direct Volume Render (DVR) of $\mid\textbf{J}\mid/\mid\textbf{B}\mid$) is shown by color bar. At $t = 120.032$s (panel (a)), nulls are spontaneously created in pairs, and the trilinear null detection technique detects them simultaneously. The generated null pair consists of two radial nulls and are shown as radial null\_1 and radial null\_2 in the figure. It can be verified by collapsing them into 2D where they appear to be akin to X-type nulls (shown in the inset of panel (a)). As the evolution continues, both radial nulls move away from each other after their generation (see panels (a)-(f)).}
\label{radial_null_pair_traced}
\end{figure}

The use of the trilinear null detection technique provides coordinates of each null along with their TD at each time step. The nulls' coordinates at two consecutive time steps are compared to segregate the new nulls from existing nulls at the previous time step. Applying the procedure, a pair of nulls (marked as radial null\_1  and null\_2) having coordinates $\{(116.045, 46.893, 5.632),  (116.047, 46.893, 5.633)\} \textnormal{ Mm}$ at $t = 120.03$ s are selected for analyses; depicted in Figure \ref{radial_null_pair_traced}, Panel (a). The eigenvalues of the Jacobian matrix $\nabla{\bf{B}}$ at each null are calculated at $t=120.03$s. The eigenvalues of each null are found to be purely real, implying that the nulls are radial nulls. The field lines are plotted near locations of the nulls for further visualization. The pair consists of radial nulls as can be verified by collapsing them into 2D where they appear to be akin to X-type nulls (shown in the inset of panel (a))---a property mentioned in \citet{1996PhPl....3..759P, 2018ApJ...860..128L, Liu_2019}. The nulls are further traced in time, indicating they are getting spatially separated with time. The cause of their separation is checked and found that it is because of Lorentz force (not shown).

\begin{figure}[!b]
\includegraphics[width=1.0\textwidth]{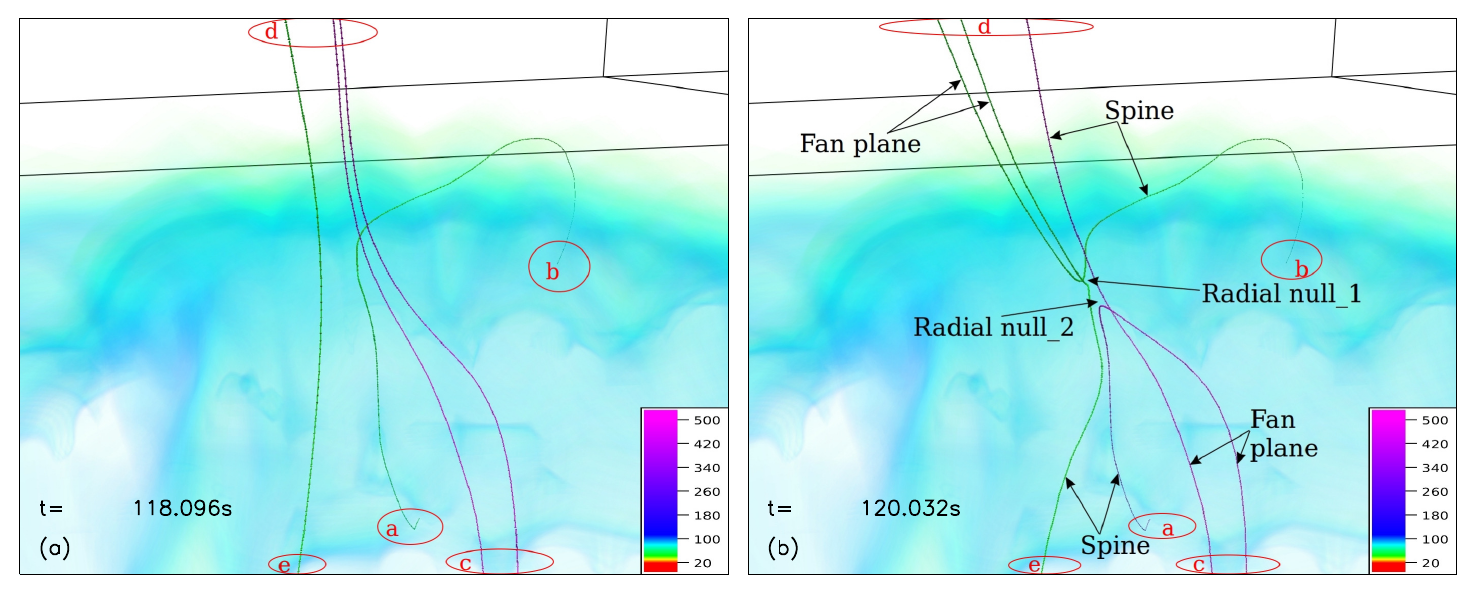}
\caption{Magnetic field lines are traced in time and advected with plasma flow. The evolution shows the creation of nulls in pairs, consisting of two radial nulls marked as radial null\_1 and radial null\_2. The enhanced current intensity is overlaid using the DVR tool in VAPOR. At $t=118.096$s, two green field lines connect from regions a to b and from e to d, while two pink field lines connect from regions c to d (panel (a)). During the evolution, one pink field line changes its connectivity from regions c to d and reconnects to regions c to a, and one green field line also changes its connectivity from regions b to a to regions b to d (panel (b)). Simultaneously, two radial nulls are created and marked by arrows in panel (b). Such changes in the connectivity of field lines represent magnetic reconnection.}
\label{reconnection_radial_null_gen}
\end{figure}

\begin{figure}[!b]
\includegraphics[width=1.0\textwidth]{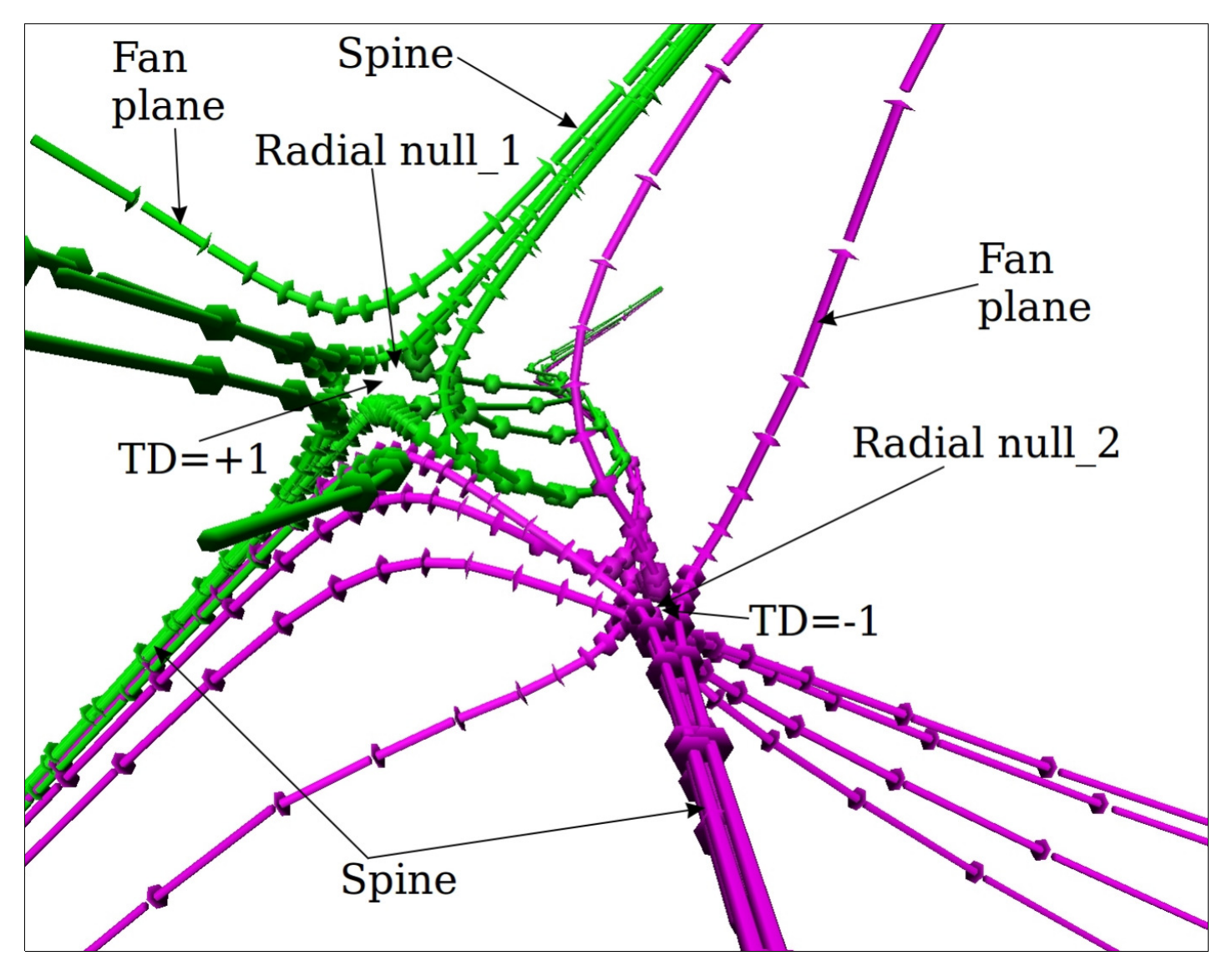}
\caption{The figure details the topological features of spontaneously generated radial nulls, at time $t=193.6 $s. Two radial nulls are generated simultaneously and are marked by arrow as radial null\_1 and radial null\_2. Green and pink field lines are drawn near the radial null\_1 and radial null\_2, respectively. The fan field lines (in green) of radial null\_1 are directed toward the null, making topological degree $+1$, while the fan field lines (in pink) of radial null\_2 are directed away from the null, making topological degree $-1$.}
\label{radial_null_topological_degree}
\end{figure}

To understand the field line dynamics responsible for the generation of the nulls,  two selected green and pink field lines at $t=118.09$s are shown in panel (a) of Fig. \ref{reconnection_radial_null_gen}. The initial points of these green and pink field lines are located away from the reconnection region and in the ideal plasma region, allowing identification of reconnection \citep{schindler1988,priest2003,Knizhnik_2022}, specifically at coordinates $(116.07, 47.05, 5.58) \textnormal{Mm}$ for the green lines and $(116.45, 46.80, 5.40) \textnormal{Mm}$ for the pink lines. The green field lines are connected from regions b to a and regions d to e, whereas pink field lines are connected from region c to region d (panel (a)). These field lines are traced in time and advected with the plasma flow. During their evolution, one of the two green field lines changes its connectivity from regions b to a and connects regions b to d. Similarly, one out of the two pink field lines also changed its connectivity from regions c to d and gets connected from regions c to a. Such changes in connectivity of field lines equate with the basal definition of  magnetic reconnection \citep{doi:https://doi.org/10.1029/GM030p0001}. Simultaneous to the reconnection, the field line topology display formation of the radial-radial null pair (marked by arrows in panel (b)), which is further corroborated by the trilinear method. Notably, such radial-radial null pair generation was absent in the earlier work by \citet{2023PhPl...30b2901M}, hereafter referred as the paper-I. For details of topology, in Fig. \ref{radial_null_topological_degree} we present magnetic field lines at the near neighbourhood of the nulls. In the figure the fan field lines (in green) of radial null\_1 are directed toward the null and are directed away from the null along the spine of the null making its topological degree $+1$. The fan field lines (in pink) of radial null\_2 are directed away from the null resulting its topological degree to be $-1$.

\begin{figure}[!b]
\includegraphics[width=1.0\textwidth]{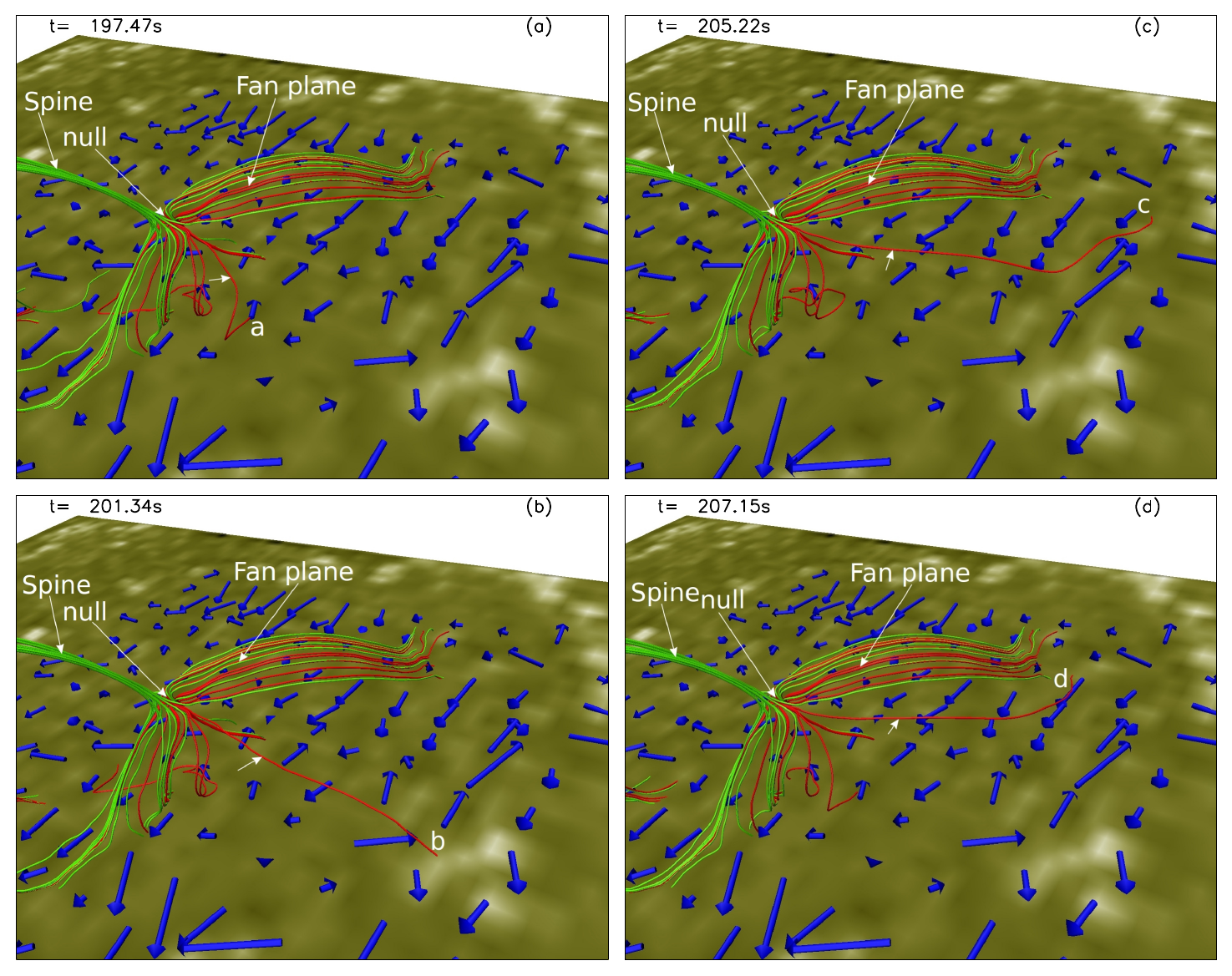}
\caption{The figure depicts the footpoint brightening in the AIA $1600$\AA channel associated with slip reconnection. The spine and the fan plane of the radial null (marked as `null') are indicated by white arrows. Two sets of field lines drawn near the radial null\_2 demonstrate the foot point brightening associated with slip reconnection. The plasma flow are plotted near the z = 0 plane and shown by blue arrows. Notably, the red field line, marked by white arrow at $t=197.47$ s is initially anchored to point `a' (panel (a)) changes its connectivity from point `a' to point `b' through slip reconnection (plasma flow direction is different than the field line motion), resulting in the associated brightening seen in panel (b). Subsequently, the red field line changes its connectivity from point `b' to point `c' and then to point `d' (refer to panels (c) and (d)). The overlaid AIA channel has dimension approximately $32.63 \textnormal{Mm} \times 63.80 \textnormal{Mm} $ in $x$ and $y$, respectively.}
\label{Slip_reconnection_brightening_1600}
\end{figure}

The data-based simulation presents a unique opportunity to check if these spontaneously generated nulls also contribute to footpoint brightening or not, another novelty of this paper. A positive outcome will strengthen the idea that such null point generation and subsequent reconnection can contribute to chromospheric/coronal heating, a concept floated by \citet{2023NatCo..14.2107C}. The reconnections at the spontaneously developed null points are expected to generate heat and accelerate particles that travel along the field lines constituting the fan and spine, resulting in footpoint brightening as they enter denser plasma region \citep{Wang_2012}. Relevantly, Figure \ref{Slip_reconnection_brightening_1600} depicts such footpoint brightening in the AIA $1600$ \AA~ channel  because of slip reconnection. For demonstration, we select a group of field lines constituting the spine and fan of null\_2 (indicated as `null'). Notably, the fan has the typical dome-shaped structure identified with 3D null in various observations \citep{Mason_2019, Mason_2021}. Blue arrows point the direction of the local plasma flow. In the figure, the red field line, marked by a white arrow at $t=197.47$ s, and initially anchored at  point `a' (panel (a)), changes its connectivity to point `b' (panel (b)) with evolution. With further evolution, the red line changes its connectivity from point `b' and reconnects to point `c' and subsequently to point `d' (refer to panels (c) and (d)). Importantly, the local plasma flow direction differs from the field line motion---a trademark of all magnetic reconnection in 3D \citep{priest2003,2006SoPh..238..347A, 2007Sci...318.1588A}. 

\begin{figure}[!b]
\includegraphics[width=0.970\textwidth]{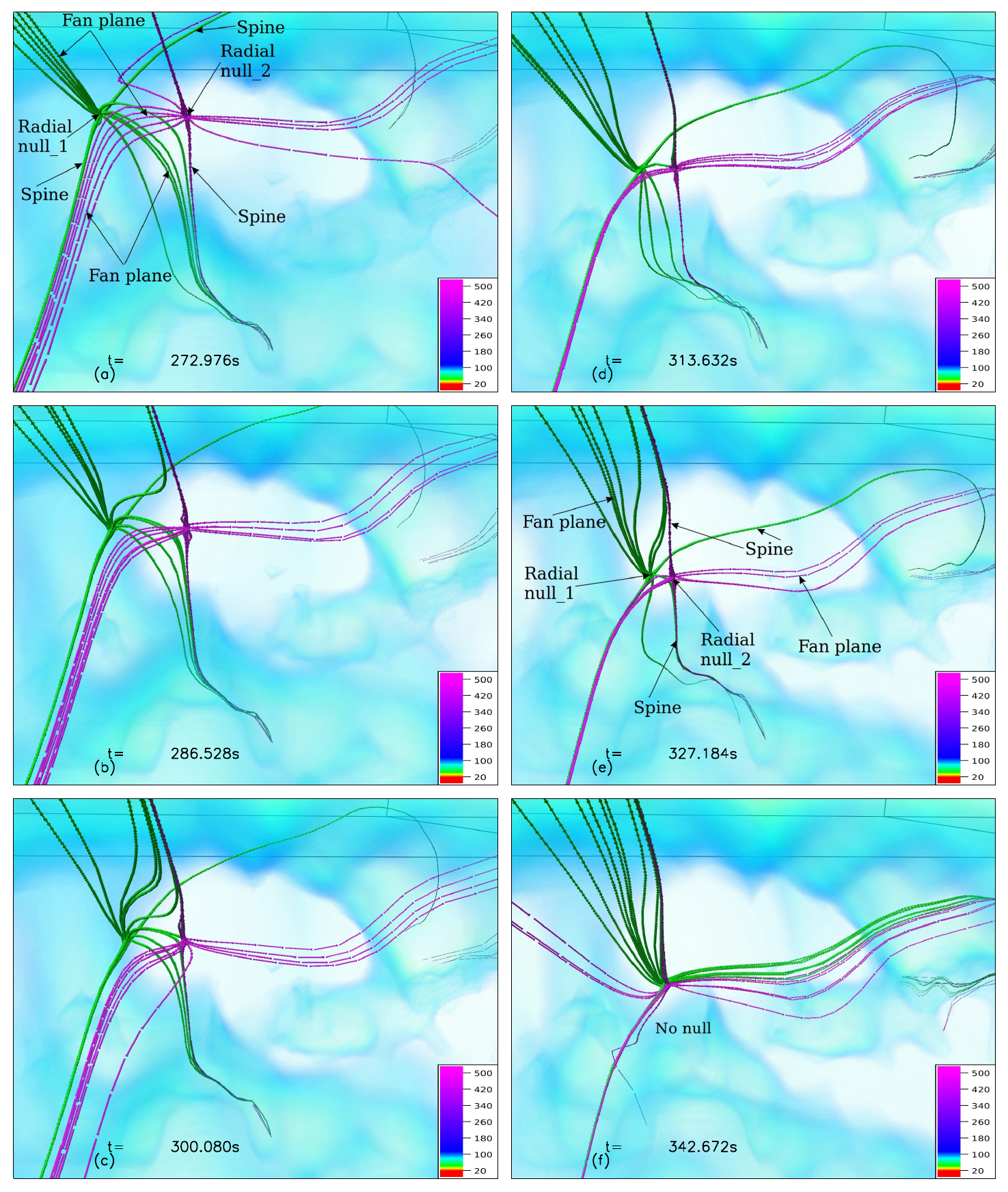}
\caption{In this figure, the evolution of radial nulls is shown by tracing and drawing field lines over time. The field lines (in green) are drawn near the radial null\_1 and field lines (in pink) are drawn at radial null\_2. The spine and fan plane of radial nulls are marked by arrows. With the evolution, radial nulls are approaching each other (panels (a)-(e)) and ultimately get annihilated at $t=342.672$ s (panel (f)).}
\label{radial_null_annihilation}
\end{figure}

\begin{figure}[!b]
\includegraphics[width=1.0\textwidth]{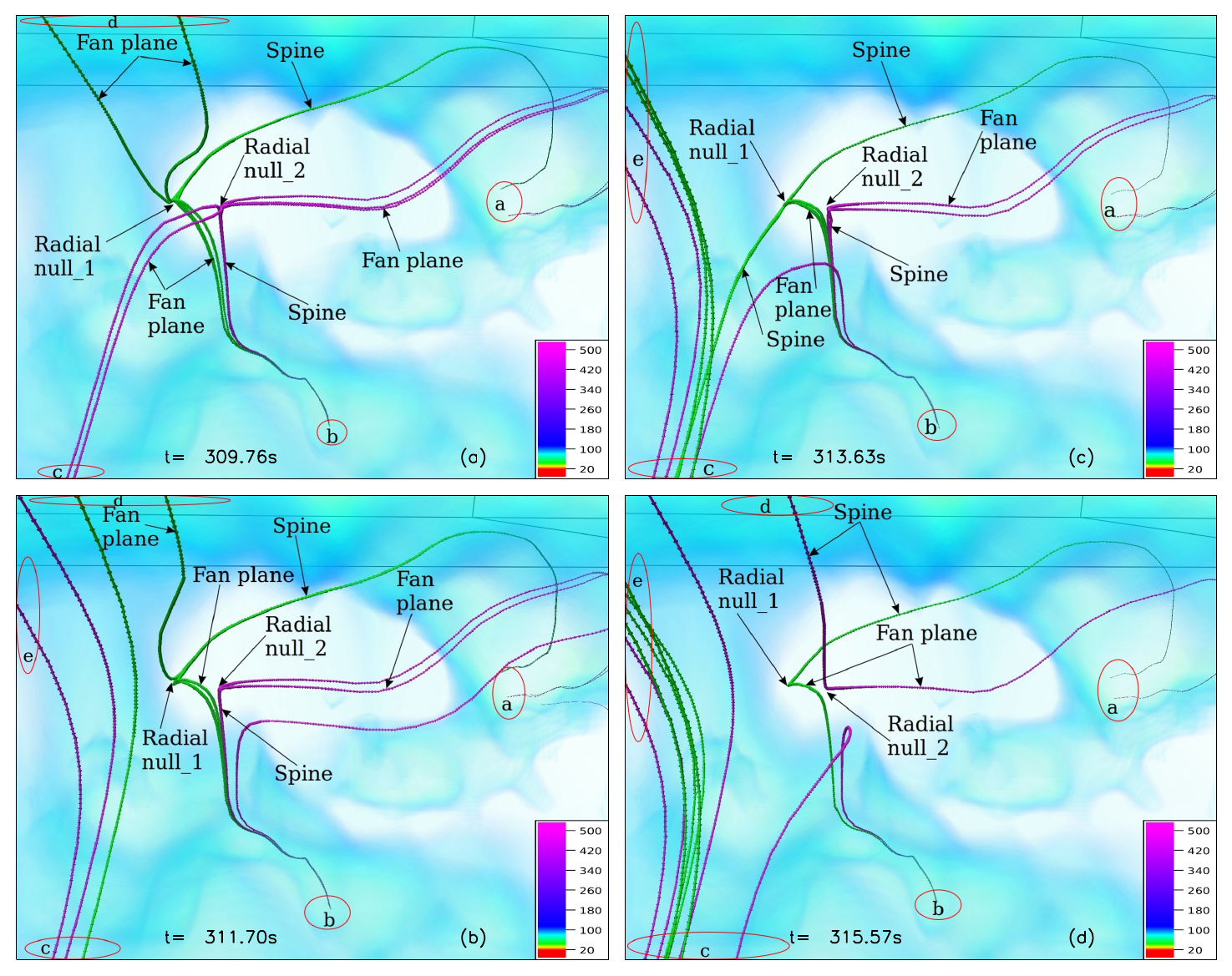}
\caption{Field lines are traced in time and advected with plasma flow. Five green and pink field lines are initially part of spine and fan plane of radial null\_1 and radial null\_2, respectively (panel (a)). With the evolution, the green and pink field lines change their connectivity and get disconnected from the nulls. Consequently, the nulls are approaching each other and ultimately annihilate each other, as shown in Figure \ref{radial_null_annihilation}.}
\label{reconnection_radial_null_annihilation}
\end{figure}

\begin{figure}[!b]
\includegraphics[width=1.0\textwidth]{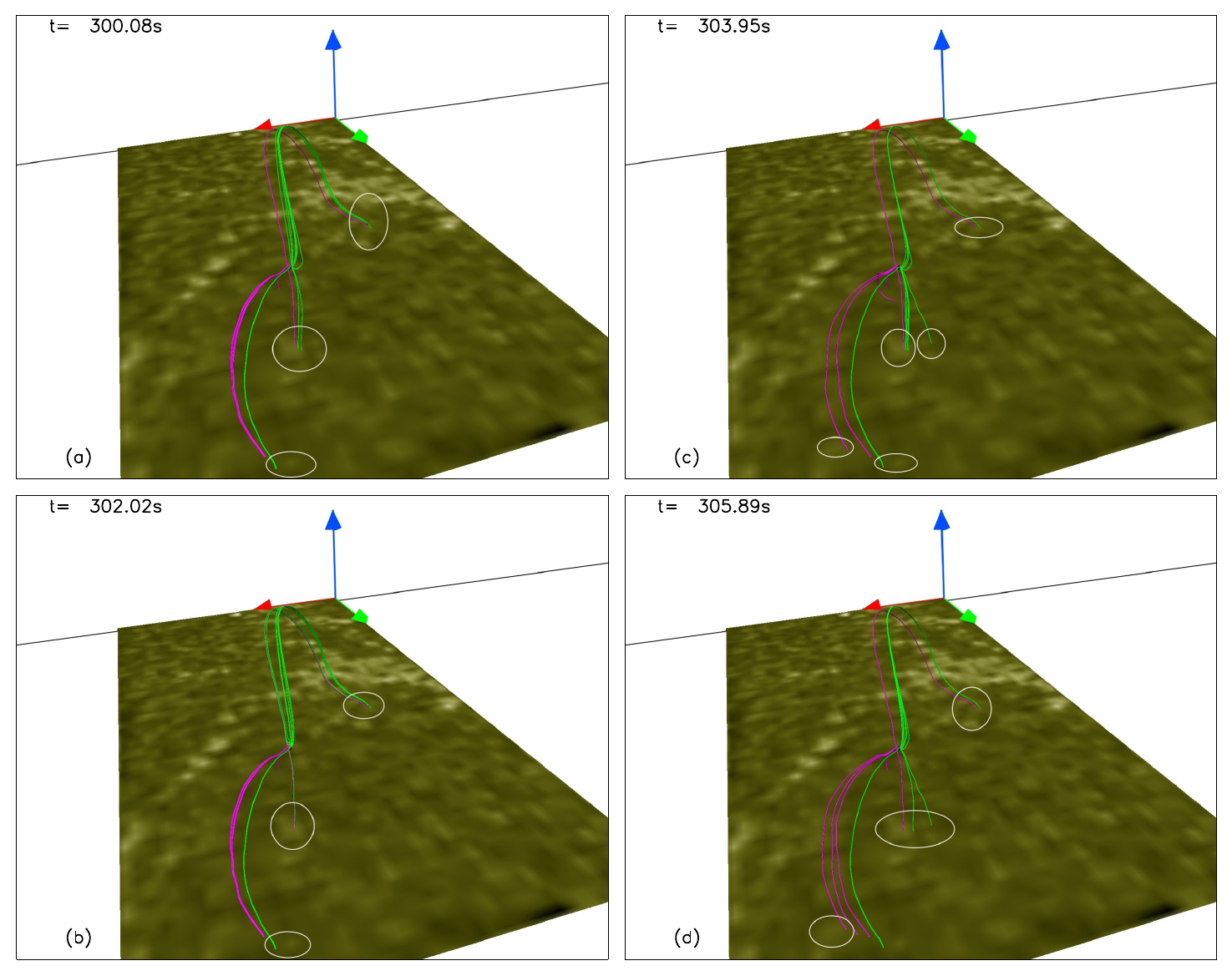}
\caption{The figure depicts foot point brightening (marked by circles)in the AIA $1600$ \AA channel associated with magnetic reconnection, which are annihilating the radial nulls. With the evolution (panels (a)-(d)), nulls are approaching each other with a change in connectivity of field lines, and the corresponding footpoint locations of field lines are co-spatial with the increased intensity in AIA $1600$ \AA filter, emulating the telltale signs of magnetic reconnection. The overlaid AIA channel has dimension approximately $32.63 \textnormal{Mm} \times 63.80 \textnormal{Mm} $ in $x$ and $y$, respectively.}
\label{brightening_radial_null_anni_1600}
\end{figure}

The nulls are further traced in time spanning $t\in \{272.97, 342.67\}$s, and field lines are drawn at nulls (Fig. \ref{radial_null_annihilation}). With the evolution, these spontaneously generated radial nulls approach each-other and ultimately annihilate at $t=342.67$s, as nulls are absent in panel (f). The result is further verified using the trilinear method. To understand the field line dynamics leading to the pair annihilation, five selected green and pink field lines are drawn in ideal region, with initial points at the locations $(116.14, 47.56, 4.37) \textnormal{Mm}$ and $(116.49, 47.55, 4.34) \textnormal{Mm}$. The selected field lines are traced over time spanning $t\in \{309.76, 315.57\}$s as they are advected with  plasma flow (Figure \ref{reconnection_radial_null_annihilation}). At $t=309.76$s, green field lines are part of the fan plane and spine of radial null\_1, anchored to the region a and connected from regions d to a and regions b to a, whereas pink field lines are part of fan plane and spine of radial null\_2, anchored in region b and connected from regions c to b and regions b to a (panel (a)). With the evolution, one green field line has changed its connectivity from regions d to a to the regions d to c, and two pink field lines have changed their connectivity from regions c to b to regions c to d and e (panel (b)). With further evolution, green field lines become the part of both the spines of radial null\_1, and a pink field line changes its connectivity from regions b to a to the regions b to c (panel (c)). In panel (d), a pink field line becomes part of the upper spine of radial null\_2, and as a result, the radial nulls approach each other. These changes in connectivity of field lines---magnetic reconnections---continue until the radial nulls annihilate each other. Figure \ref{brightening_radial_null_anni_1600} depicts brightening in AIA $1600$ \AA channel, co-spatial  with footpoints of the reconnecting field lines (marked by circles). 

\subsection{Radial-spiral-pair-1} \label{subsec:radial-spiral-pair-1}

\begin{figure}[!b]
\includegraphics[width=1.0\textwidth]{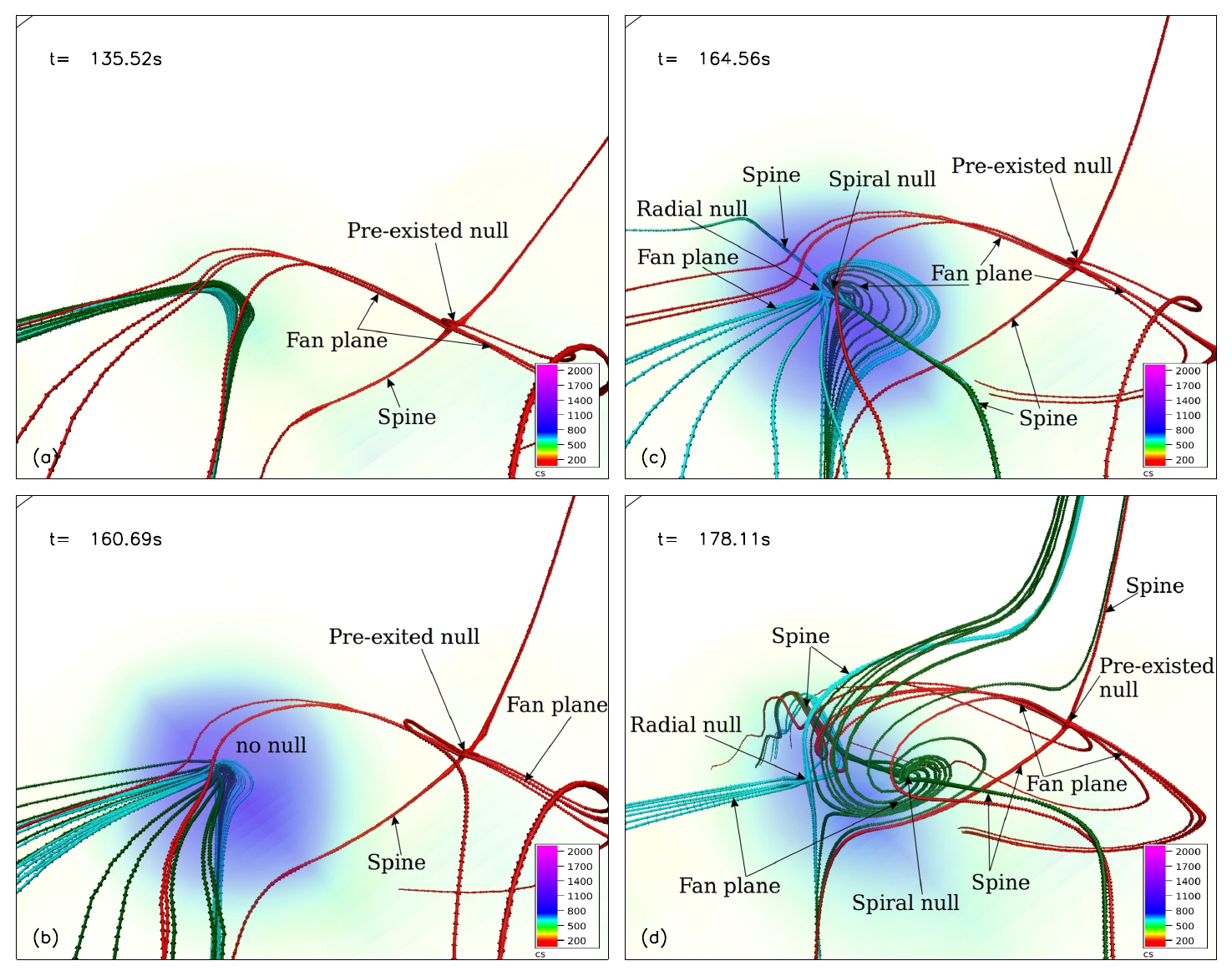}
\caption{The panels of figure illustrate the generation of nulls in a pair near a pre-existing 3D null. Magnetic field lines (in red) are drawn near the pre-existing null (a null already present at $t = 135.52$s), while sky-blue and green field lines are included to facilitate the generation of nulls at a later time (panel (a)). With the evolution, the sky-blue and green field lines develop an elbow shape at around $t = 160.69$ s, and an enhancement in current intensity (identified by the Direct Volume Render of $\mid\textbf{J}\mid/\mid\textbf{B}\mid$), marked as `cs', is seen accordingly (panel (b)). In panel (c), a pair of nulls consisting of a radial and a spiral null is generated at $t= 164.56$ s. Panels (c)-(d), spanning $t\in{164.56, 178.11}$ s, depict the tracing of nulls and the plotting of field lines. As the evolution progresses, the radial and spiral nulls move away from each other after their generation, whereas the spiral null of the generated pair approaches the pre-existing null}
\label{culprit_null_generation}
\end{figure}

\begin{figure}[!b]
\includegraphics[width=1.0\textwidth]{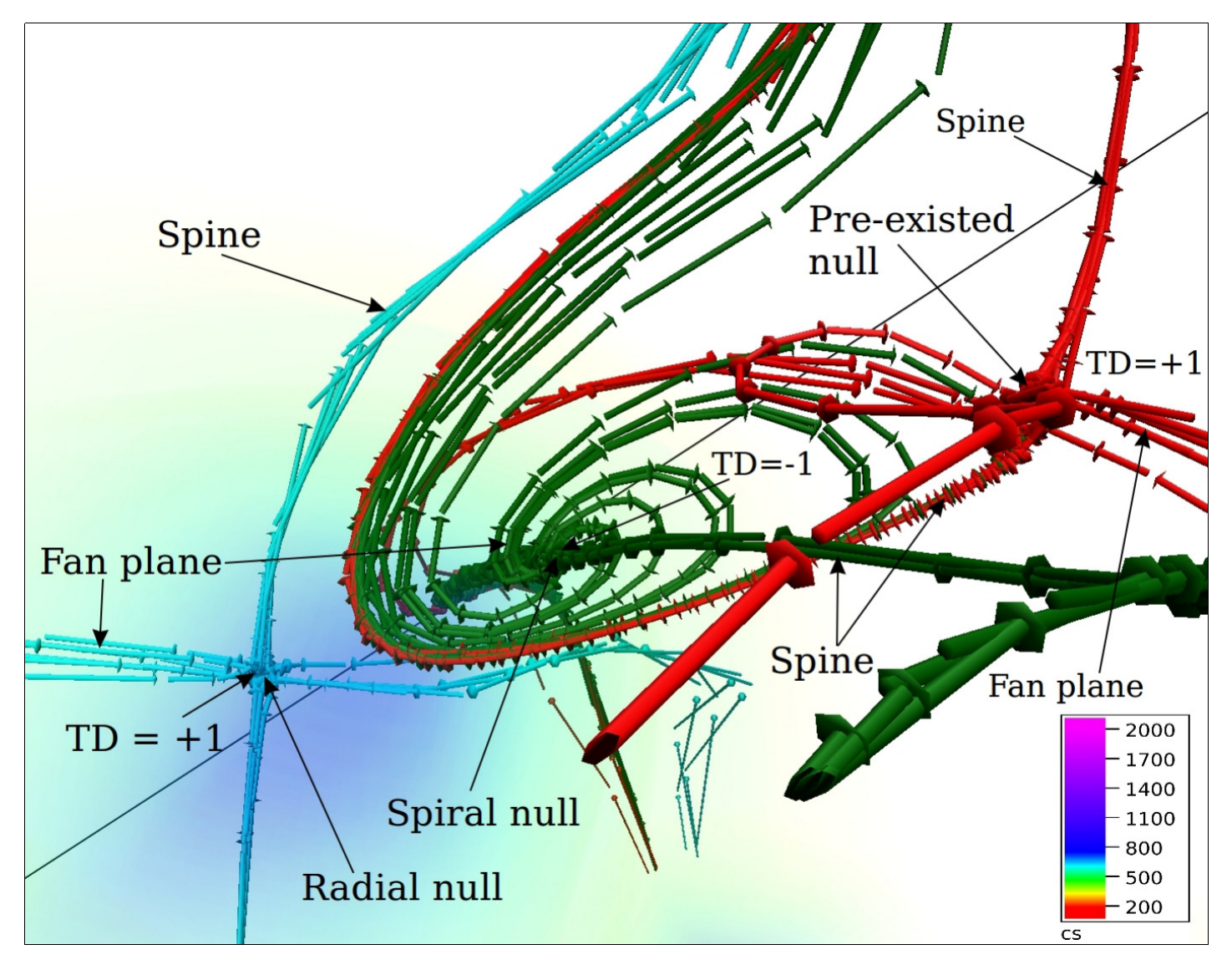}
\caption{The figure illustrates the details of a pre-existing null along with a spontaneously generated null pair, which consists of a radial null and a spiral null at $t=178.11 $s. The field lines (in red) are drawn near the location of pre-existing null. The fan field lines are directed toward the null, resulting in a topological degree of $+1$. Field lines drawn near the radial null (in sky-blue) and those of the spiral null (in green) are also shown. The spine and fan planes, along with the topological degrees of both nulls are marked in the figure. The direction of the fan field lines of the radial null is toward the null point, and the spine field lines are directed away from the null point, resulting in a topological degree of $+1$. On the other hand, the spine field lines of the spiral null are directed toward the null point in the fan plane and away from the null point, resulting in a topological degree of $-1$. The net topological degree of this generated pair is zero, and the spiral null gets annihilated with the pre-existing null (in a pair). Therefore, the conservation of the net topological degree is self-explanatory.}
\label{topological_degree_culprit_null}
\end{figure}

Additional to the radial-radial pair generation, simulation also shows generation of other pairs. This subsection focuses on the generation and annihilation of a radial-spiral null pair. The generation and annihilation mechanism being similar to the ones presented in paper-I, here only a brief description is provided. The generation can be visualized by tracking magnetic structure in the immediate vicinity of the null pair, shown in Fig. \ref{culprit_null_generation}. The uniqueness in this case is the role of a pre-existing null in annihilating the pair. For visualization, magnetic field lines (in red) are drawn near the pre-existing null while sky-blue and green field lines are drawn to facilitate demonstration of the null generations (panel (a) of Fig. \ref{culprit_null_generation}) and their evolution (subsequent panels). With evolution, sky-blue and green field lines get elbow shaped at around $t = 160.69$s, and an enhancement in current intensity (identified in DVR of $\mid\textbf{J}\mid/\mid\textbf{B}\mid$) is seen accordingly (panel (b)). At $t= 164.56$s nulls in a pair comprising of a radial and spiral null get spontaneously created (panel (c)). Across panel (c) and (d) nulls are traced in time and field lines are drawn at their near neighbourhood, depicting an increasing separation between the radial and spiral nulls with time. The pair generation is due to reconnection, confirmed by advecting participating field lines as in the previous case (not shown here). The topological degree of spontaneously generated nulls together with pre-existing null are depicted in Figure \ref{topological_degree_culprit_null}. The fan field lines (in red) are drawn near the location of pre-existing null and are directed toward the null rendering its topological degree to be $+1$. The direction of the fan field lines (in sky-blue) of radial null is toward the null point resulting a topological degree of $+1$, whereas the fan field lines (in green) of spiral null are directed away from the null point resulting in a topological degree of $-1$. The net topological degree of the generated pair is zero, indicating its preservation during formation of the pair.  

\begin{figure}[!b]
\includegraphics[width=1.0\textwidth]{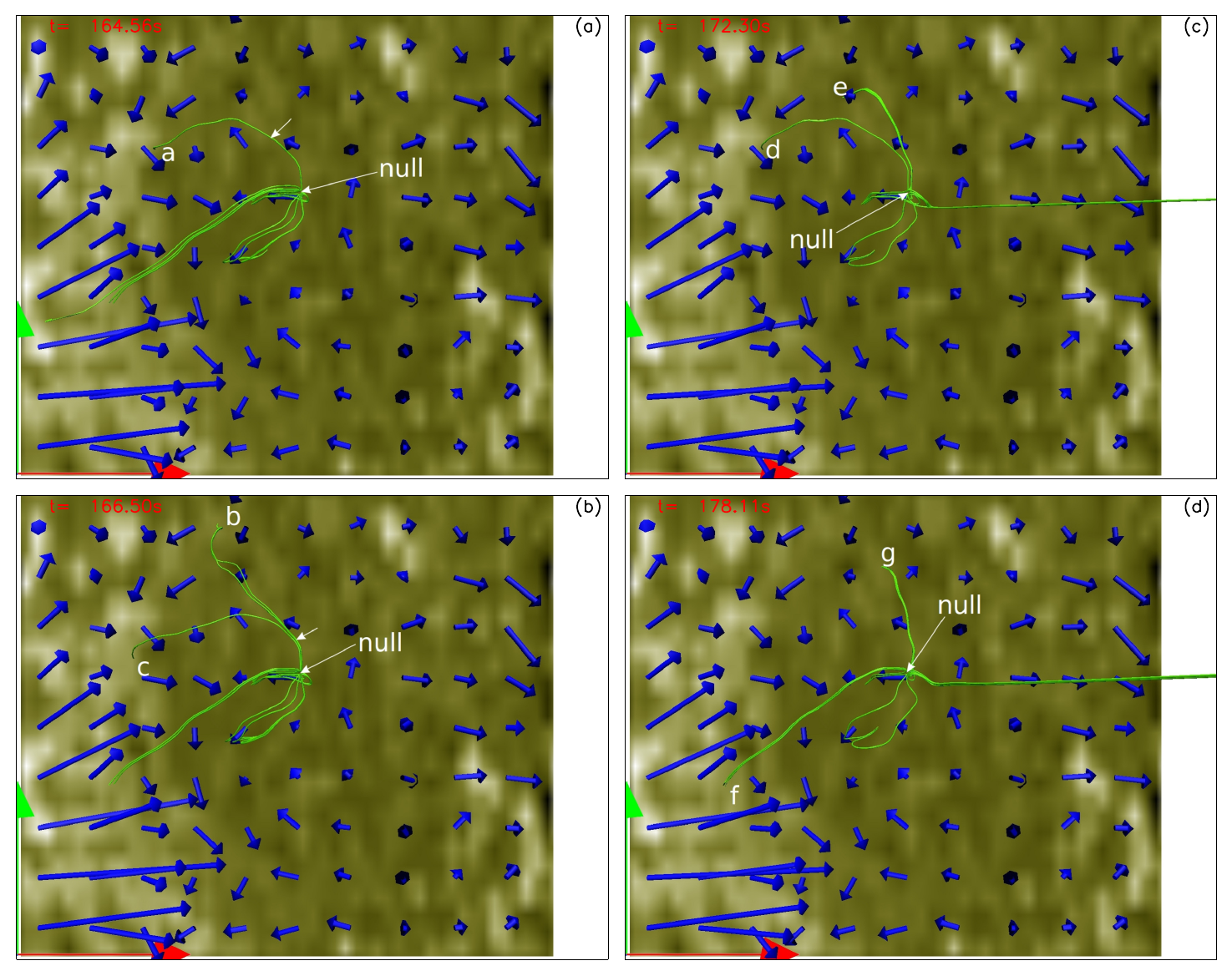}
\caption{Panels depict the footpoint brightening corresponding to the slip reconnection of fan field lines of the radial null of the spiral-radial null pair-1. The radial null is marked as ``null" and local plasma flow shown by blue arrows are plotted near the z = 0 plane. Initially, at $t=164.35$ s, the green field line indicated by the white arrow is anchored to point a (panel (a)). With the evolution, the footpoints of the green field lines are changing their connectivity to points b and c (panel (b)) and subsequently to points d, e, f, and g due to slip reconnection. The overlaid AIA channel has dimension approximately $21.75 \textnormal{Mm} \times 21.75 \textnormal{Mm} $ in $x$ and $y$, respectively.}
\label{spiral_slip_rec}
\end{figure}

The expected slip reconnection by the fan field lines \citep{2013ApJ...774..154P} and the corresponding brightening in AIA 1600 \AA~ channel can be inferred from Fig. \ref{spiral_slip_rec}. Panels depict the footpoint brightening corresponding to the slip reconnection of fan field lines of the radial null of the spiral-radial null pair-1. The radial null is marked as ``null" and local plasma flow is shown by blue arrows. Initially, at $t=164.35$ s, the green field line indicated by the white arrow is anchored to point a (panel (a)). With the evolution, the footpoints of the green field lines are changing their connectivity to points b and c (panel (b)) and subsequently to points d, e, f, and g due to slip reconnection. Importantly, the local plasma flow direction differs from the field line motion---a trademark of slip reconnection.

\begin{figure}[!b]
\includegraphics[width=1.0\textwidth]{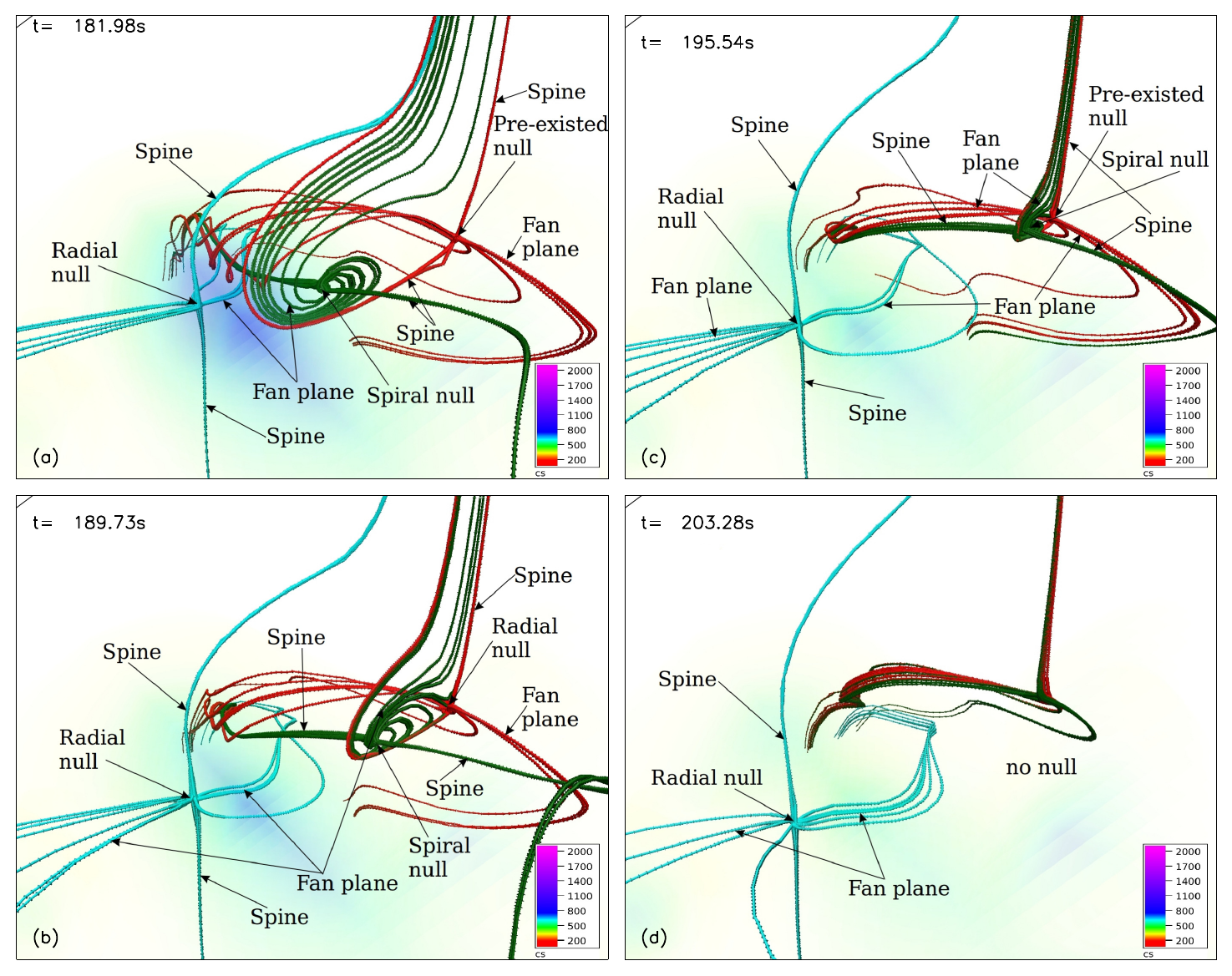}
\caption{Nulls are traced over time, and field lines are drawn near the locations of the nulls. Panels (a)-(d) of this figure illustrate the annihilation of a spiral null and a pre-existing null. The spine and fan planes of the radial, spiral, and pre-existing nulls are indicated by the arrows.}
\label{culprit_null_annihilation}
\end{figure}

\begin{figure}[!b]
\includegraphics[width=1.0\textwidth]{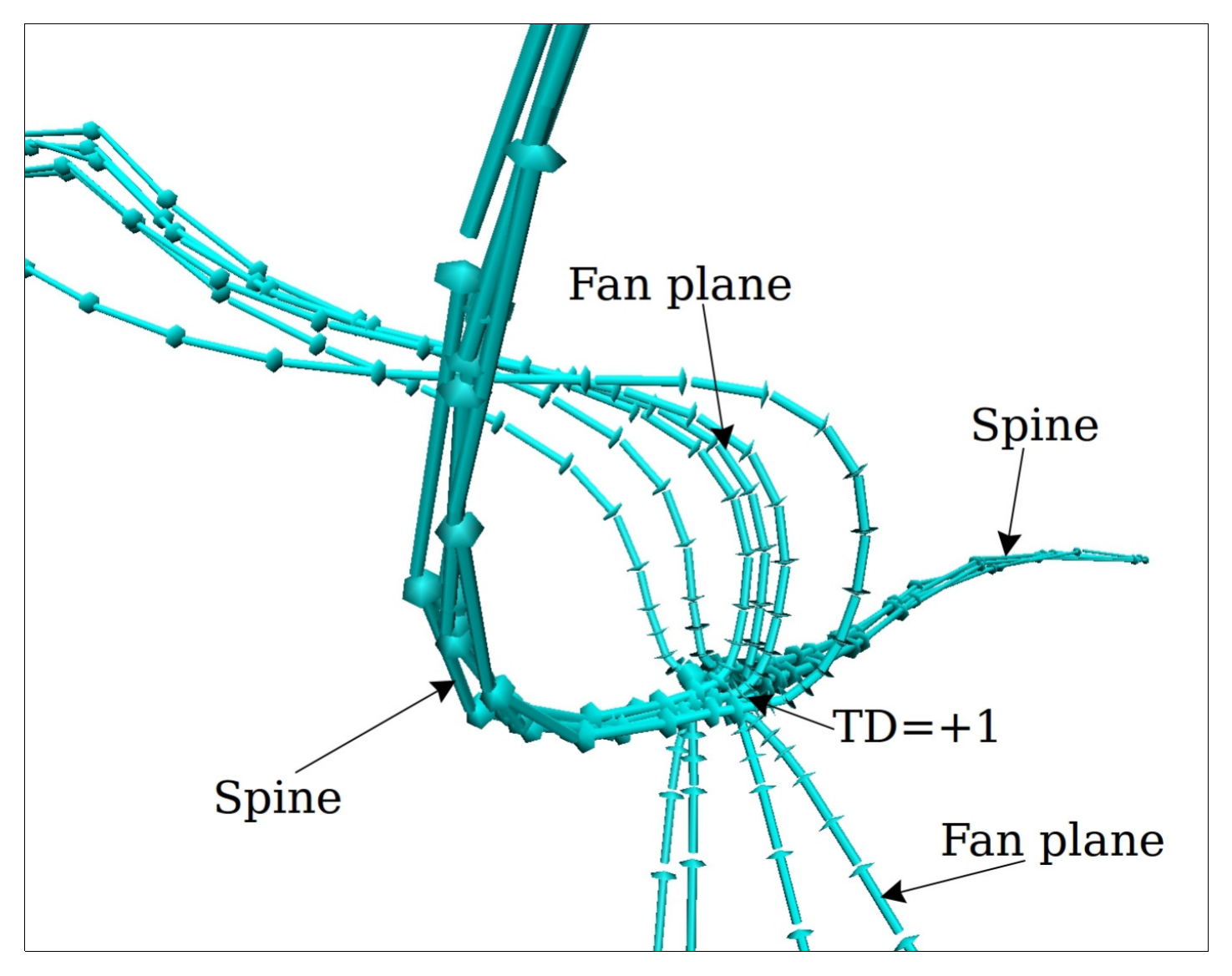}
\caption{The figure illustrates the details of the radial null $t=203.28 $s, left in the domain after the annihilation of the generated spiral null with the pre-existing null. The fan field lines of the radial null point toward the null point, while the spine field lines are directed away from the null point, resulting in a topological degree of $+1$.}
\label{td_radial_null}
\end{figure}

Figure \ref{culprit_null_annihilation} illustrates the evolution of spontaneously generated nulls together with the pre-existing null. The nulls are traced in time, and field lines are drawn near their locations. Across panels (a)-(d), the spiral null is receding away from the radial null and approaches the pre-existing null. At $t=203.28$s, the spiral and pre-existing nulls are annihilated and correspondingly only the spontaneously generated radial null is present in panel (d)---independently verified using the trilinear method.  The cause of the null annihilation has been investigated by advecting the relevant field lines (not shown here) and found to be due to magnetic reconnection. The field line topology of the remaining radial null is shown in  Figure \ref{td_radial_null}. Its topological degree is $+1$ as the fan field lines are directed toward the null, in conformity with conservation of topological degree.

\subsection{radial-spiral-pair-2} \label{subsec:decent}

\begin{figure}[!b]
\includegraphics[width=0.90\textwidth]{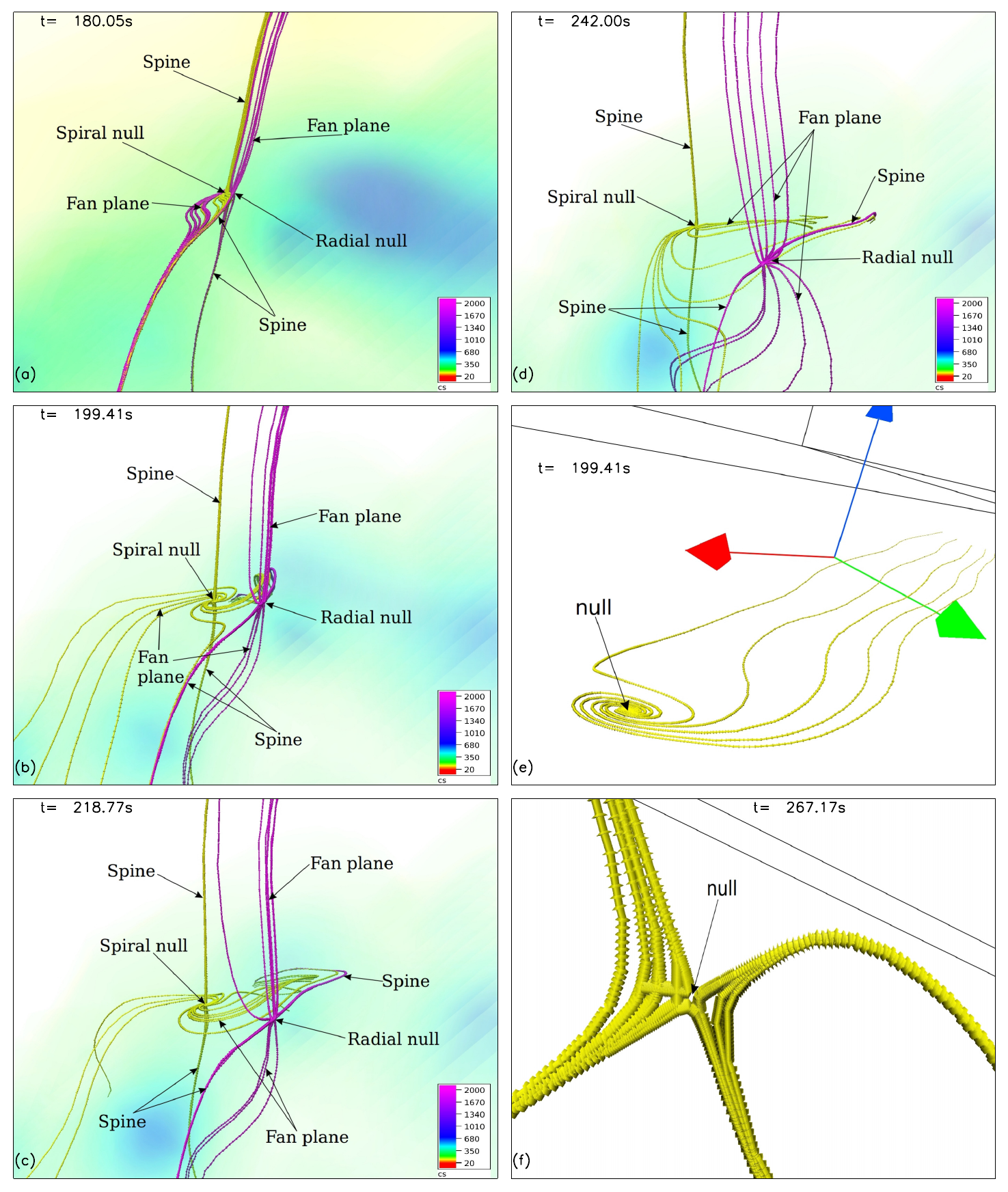}
\caption{The snapshots of the field lines represent the evolution of nulls as they are traced over time. At $t=180.05$ seconds, it is the first instance when nulls in a pair first appear using the trilinear detection technique, and field lines are drawn near their locations (panel (a)). These spontaneously generated nulls are named as radial-spiral-pair-2 consists of a spiral null (in yellow) and a radial null (in pink). As the evolution progresses, the spiral and radial nulls are moving away from each other (panels (a)-(d)). The spiral null looses its spirality and gets converted into a radial null, The conversion from spiral to radial null can be verified by collapsing the null's structure in 2D, where a spiral null will appear as an ``O" type and a radial null will appear as an ``X" type null. Panels (e) and (f) depict the 2D projections of the spiral null and converted radial null at $t=199.41$s and $t=267.17$s, respectively. This illustration shows a similar conversion from ``O" to ``X" type.}
\label{tracing_of_decent_nulls}
\end{figure}

\begin{figure}[!b]
\includegraphics[width=1.0\textwidth]{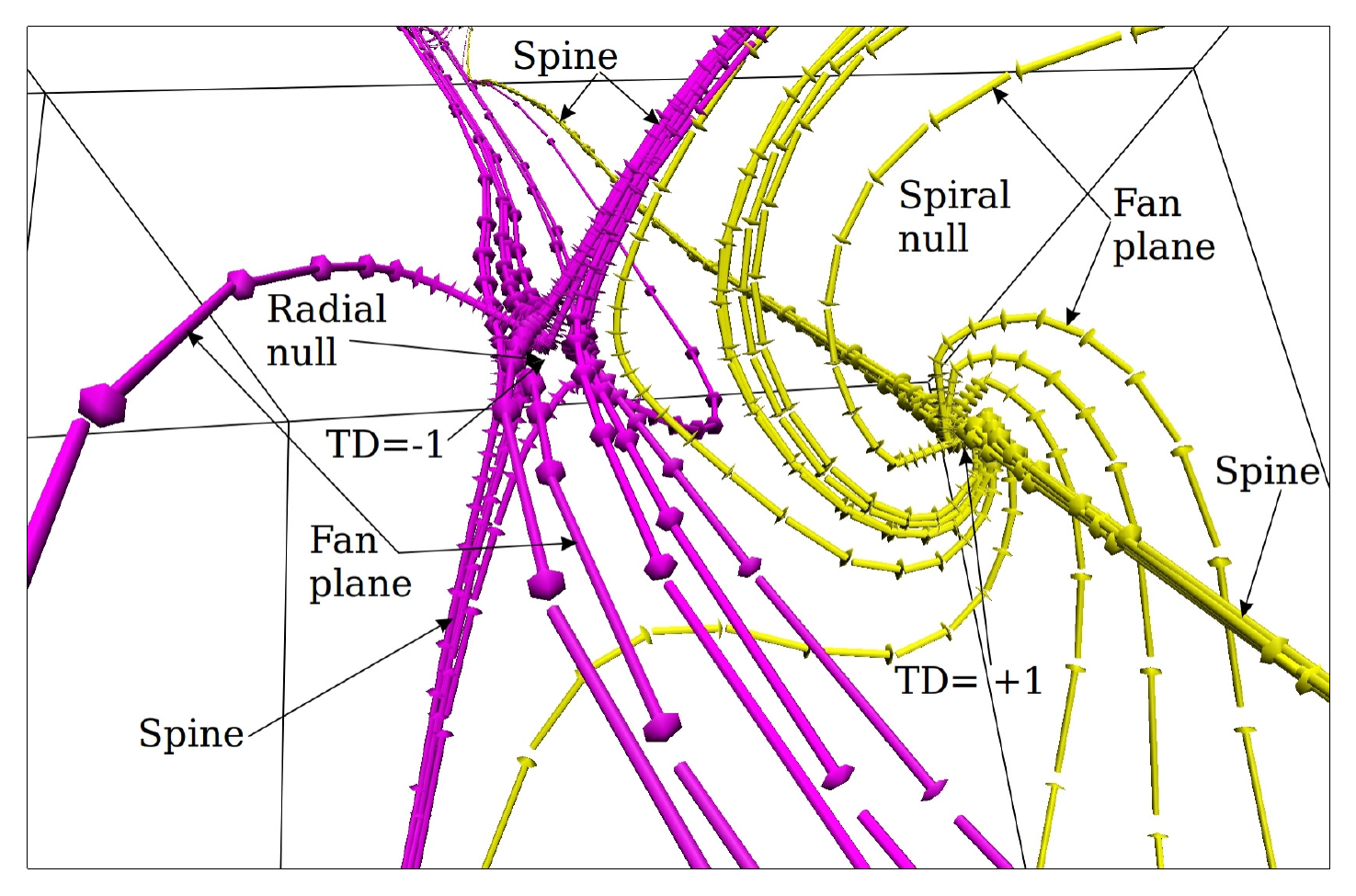}
\caption{The figure illustrates the topological details of spontaneously generated radial-spiral-pair-2 nulls at $t=199.41 $s. These nulls are generated in a pair and consist of a spiral null (in yellow) and a radial null (in pink). The spine field lines (in pink) of the radial null are directed toward the null point, resulting in a topological degree of $-1$, while the fan field lines (in yellow) of the spiral null are directed toward the null point, making a topological degree $+1$. The net topological degree of this local system is zero, and hence, the overall topological degree of the system remains unaffected by the generation of these new nulls.}
\label{td_decent_null}
\end{figure}

This subsection emphasizes the generation and annihilation of the radial-spiral-pair-2, where a radial-spiral null pair spontaneously generates, move away from each other after the generation (panels (a)-(d) of Figure \ref{tracing_of_decent_nulls}) and subsequently annihilates. Magnetic reconnection is once again the cause of the pair generation and annihilation---nevertheless, the uniqueness here is conversion of the spiral null into a radial null which later annihilates with another spiral null of a newly generated radial-spiral null pair. To highlight this uniqueness, eigenvalues of the Jacobian matrix $\nabla{\bf{B}}$ at the spiral null is calculated during its evolution. The imaginary part of the eigenvalue is zero at $t=267.17$s while being non-zero earlier than that, implying the transition from spiral to radial. For further visualization, field lines are plotted near locations of the nulls. In 2D, a spiral null will appear as an ``O" type and a radial null will appear as an ``X" type null \citep{1996PhPl....3..759P, 2018ApJ...860..128L, Liu_2019}. Panels (e) and (f) depict such projections of the spiral null at $t=199.41$s and $t=267.17$s, illustrates  similar conversion from ``O" to ``X" type. The spine, fan plane and the topological degree of nulls at $t=199.41 $s are shown in Figure \ref{td_decent_null}. The fan field lines (in yellow) of the spiral null are directed toward the null point, making its topological degree $+1$, whereas the spine field lines (in pink) of radial null are directed toward the null, resulting in a topological degree of $-1$. Consequently, the net topological degree of this pair is zero and the generation is in congruence with the conservation of topological degree. 
 
\begin{figure}[!b]
\includegraphics[width=1.0\textwidth]{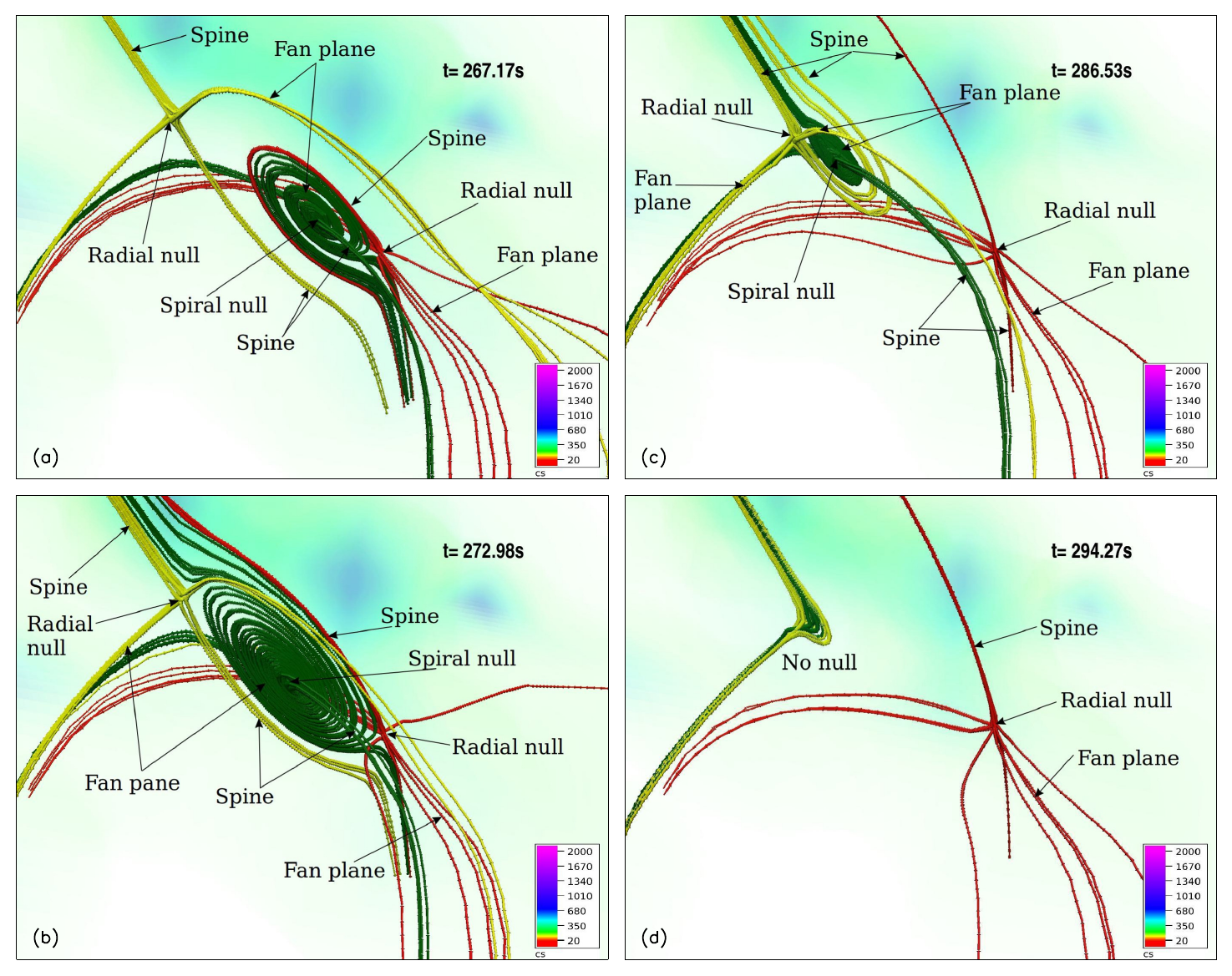}
\caption{The figure illustrates the evolution of nulls over time. These nulls are tracked as they evolve, and field lines are drawn at their locations. Yellow, green, and red field lines are drawn near the radial null of the radial-spiral-pair-2 null pair, the spiral null, and the radial null of a spontaneously generated new null pair, respectively. The spontaneously generated new nulls are first detected by the trilinear null detection technique at $t=267.17$ seconds, and the corresponding structure is shown in panel (a). As the evolution continues, the spiral and radial nulls of the newly generated pair move away from each other, while the radial null of the radial-spiral-pair-1 pair and the spiral null of the new generated null approach each other simultaneously (panels (a)-(c)), ultimately resulting in annihilation around $t=294.27$ s (marked by no null in panel (d)).
The pairwise annihilation does not affect the net topological degree of the system. Consequently, one radial null, with a topological degree of $+1$, is left in the system.}
\label{Annihilation_of_decent_null}
\end{figure}

\begin{figure}[!b]
\includegraphics[width=1.0\textwidth]{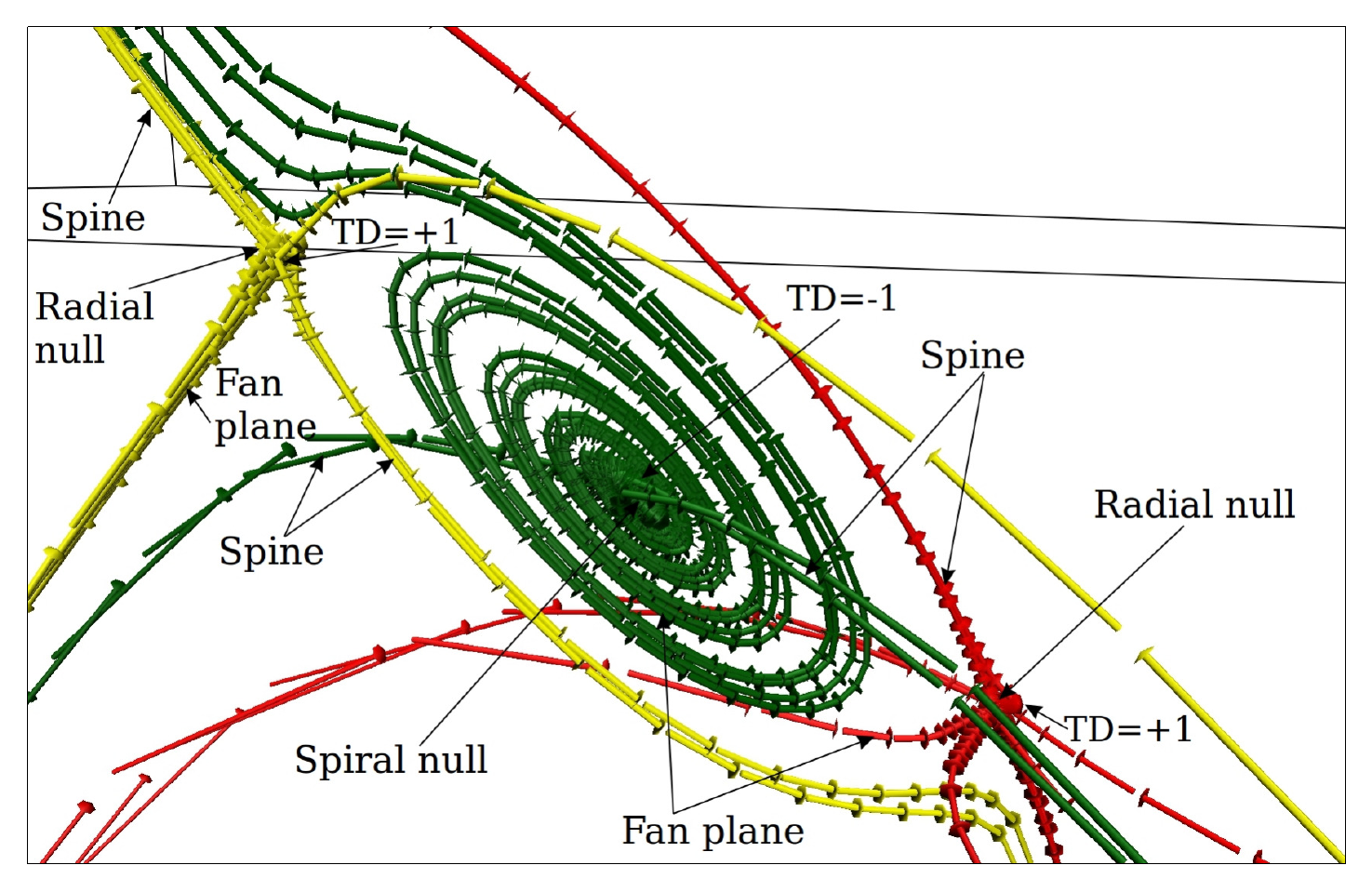}
\caption{The figure illustrates the topological details of spontaneously generated new nulls near the radial null of the radial-spiral-pair-2 pair at $t=274.91 $s. These new nulls are generated in a pair and are marked by the arrows. The fan field lines (in yellow) of the radial null of the radial-spiral-pair-1 pair are directed toward the null point, resulting in a topological degree of $+1$. Meanwhile, the fan field lines (in green) of the spiral null of the spontaneously generated new pair are directed away from the null point, resulting in a topological degree of $-1$. Lastly, the fan field lines (in red) of the radial null of the spontaneously generated pair are directed toward the null point, making the topological degree $+1$. Hence, spontaneous generation occurs in a pair, and the annihilation of the radial null of the radial-spiral-pair-1 pair is with the spiral null of the generated pair. Therefore, the net topological degree of the system remains unaffected by the generation and annihilation of 3D nulls.}
\label{td_of_new_null_alongwith_radial_null}
\end{figure}

\begin{figure}[!b]
\includegraphics[width=1.0\textwidth]{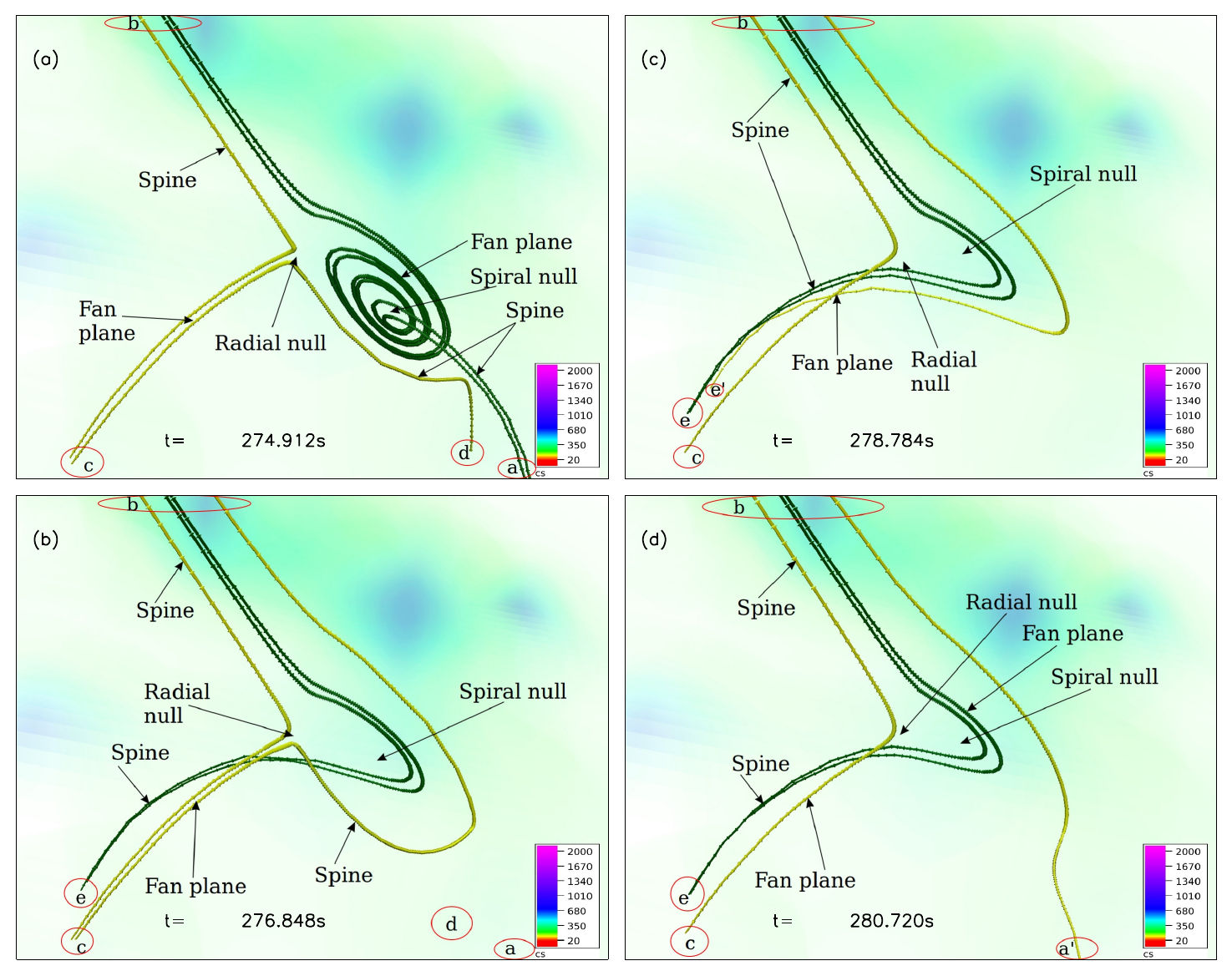}
\caption{Magnetic field lines are traced over a time span of $t\in {274.912, 280.720}$ s and advected with plasma flow. This time span is selected to investigate the dynamics of field lines responsible for annihilation. Four selected field lines, two green and two yellow, are drawn to illustrate magnetic reconnection. At $t = 274.912$ seconds, the two green field lines are part of the spine and fan plane of the spiral null, connecting from region b to region a (panel (a)), while the two yellow field lines are part of the upper and lower spine along with the fan plane of the radial null, connecting regions c to d and b. As the evolution progresses, the green field lines change their connectivity from region b to region a to region b to e. Similarly, one yellow field line also changes its connectivity from region c to d to region c to b (panel (b)). With further evolution, one yellow field line changes its connectivity from region c to b to region e$^\prime$ (panel (c)), and then from region e$^\prime$ to region b to region a$^\prime$ to region b. These changes in connectivity occur through magnetic reconnection, ultimately resulting in the annihilation of nulls in a pair.}
\label{reconnection_annihilation_decent_null}
\end{figure}

Fig.~\ref{Annihilation_of_decent_null} demonstrates the annihilation of the converted radial null along with the spiral null of the newly generated radial-spiral null pair where the green field lines corresponds to the spiral null whereas the red field lines belong to the radial null. Across panels (a)-(c), spanning the time $t\in \{267.17, 286.53\}$s, the radial null of the radial-spiral-pair-2 and the spiral null of generated pair approach each other and ultimately annihilate at $t=294.27$s. Post annihilation, a radial null remains in the system (panel (d)) and the conservation of net topological degree is self explanatory. The topological degree, spine, and fan plane of the radial null and the spontaneously generated nulls are depicted in Fig. \ref{td_of_new_null_alongwith_radial_null}. For completeness, Figure \ref{reconnection_annihilation_decent_null} illustrates changes in field line connectivity during the annihilation. Two selected yellow and green field lines are traced in time and advected with the plasma flow, preserving their uniqueness. At $t=274.91$s, initially, yellow field lines belong to the spine and fan plane of the converted radial null, while green field lines are part of the spiral null (panel (a)). With the evolution, one yellow field line changes its connectivity from regions c to d to regions c to b and becomes part of the fan plane of the spiral null. The two green field lines are changing their connectivity from regions b to a to regions b to e (panel (b)). Subsequently, during the evolution, yellow field line further changes its connectivity from regions b to c and gets connected from the regions b to e$^\prime$ (panel (c)) and then from regions b to e$^\prime$ to regions b to a$^\prime$. Slip reconnection and brightening in AIA 1600\AA~  channel similar to the other pairs have also been found, but are not shown here to avoid repetitions.

\section{Summary} \label{sec:summary}
The article establishes spontaneous generation and annihilation of magnetic nulls in the solar atmosphere through a combination of data-based Implicit Large Eddy simulation, a modified trilinear null detection technique and observation of collocated footpoint brightening caused by these generated nulls through slip reconnection. For the purpose,  magnetic field lines corresponding to Active Region AR11977 hosting a C-class flare on 17 February, 2014 is selected. The magnetic field line configuration just before the flare has been constructed with a non force-free extrapolation of the photospheric magnetic field obtained from the HMI, on board the SDO. The corresponding finite Lorentz force initiates the dynamics, without the requirement of any prescribed flow. Since this Lorentz force depends on the photospheric magnetic field, is inherent to the field line complexity of the active region and hence, the subsequent null generation can be envisaged to be spontaneous in its true sense. The modified trilinear method sheds light on the topological degree of a null, additional to finding location of a null--- as its predecessor. This new information makes it convenient to identify members of a null pair and trace them in time. The physics of reconnection being scale independent, these spontaneously generated nulls are further expected to exhibit signatures analogous to observed 3D null assisted solar flares, particularly the footpoint brightening because of slip reconnection \citep{2009ApJ...700..559M,2013ApJ...774..154P}. Accordingly, it is attempted to look for the brightening and slip reconnections associated with these spontaneously generated null points. Overall, the article establishes spontaneous generation of 3D nulls in the solar atmosphere in a realistic scenario by carrying out a data-based numerical simulation of the solar atmosphere, and explores their observational implications.

For the carried out ILES, the magnetofluid is idealized to be thermodynamically inactive, implicitly dissipative, incompressible, viscid and having an effective fluid Reynolds number 5000 which is 5 times smaller than the coronal value. The initial magnetic field is provided from the non-FFF extrapolation and initial plasma flow is set to zero. The non-zero Lorentz force pushes the plasma to generate initial dynamics. The simulation covers a fraction of the rising phase of the flare which is expected to be reconnection dominated.

The  magnetic field initially contains over a thousand null points, with this number varying throughout the simulation as pair creation and annihilation processes occur. As is expected from previous works, as the system relaxes towards an equilibrium (due to the non-zero viscous damping) the overall number of nulls decreases. Three particular null creation/annihilation events are selected for detailed study, and are representative of the many further events that occur within the domain. One of these identifies that two radial nulls are spontaneously created and move away from each other after their generation. The result is novel, as the creation of radial nulls in pairs has not been studied in earlier works. The underlying cause of their generation is magnetic reconnection, identified by advecting the magnetic field lines while selecting seed points away from reconnection sites. The spontaneously generated radial null exhibits the typical dome-shaped structure of fan field lines, along with footpoint brightening in 1600 \AA~ channel of AIA associated with slip reconnection. With further evolution, both radial nulls approach each other and spontaneously annihilate, the underlying cause being identified as magnetic reconnection. The simulation also shows the spontaneous generation of a pair of nulls, consisting of a spiral and a radial null with topological degrees $-1$ and $+1$, respectively. This pair is spontaneously created near a pre-existing 3D null with a topological degree of $+1$. The spontaneously generated radial null also exhibits the typical footpoint brightening corresponding to slip reconnection. With evolution, the spiral null moves away from the radial null while simultaneously approaching the pre-existing null, ultimately resulting in their annihilation. Here, the unique aspect of the finding is the spontaneous annihilation of the generated spiral null with a pre-existing radial null, which has not been explored in previous studies. During the evolution, another pair of nulls, consisting of a spiral null and a radial null with topological degrees of $+1$ and $-1$, respectively, gets spontaneously created. The spiral and radial nulls move away from each other after their generation, and the spiral null loses its spirality as it evolves and transforms into a radial null, making this pair unique for study. Another uniqueness here is that the annihilation of the converted radial null involves another spiral null from a newly generated radial-spiral null pair. The newly generated spiral and radial null have topological degrees of $-1$ and $+1$, respectively. Magnetic reconnection is being identified as the underlying cause of the generation and annihilation of 3D nulls through the process of advecting the field lines in all pairs. The generation and annihilation of 3D nulls in all above pairs maintain conservation of net topological degree, contributing to the credibility of the simulation. 

Magnetic reconnection, identified as the underlying cause of the spontaneous generation and annihilation of nulls, is noteworthy. The spontaneously generated nulls also exhibit slip-reconnection, a phenomenon typically observed in a 3D null of the solar atmosphere. The findings shed light on the underlying magnetic field line dynamics governing 3D null generation, annihilation, and their evolution. In each selected case, the field lines in the vicinity of the null are rooted in the photosphere near brightenings. Such brightenings may be produced by the impact of non-thermal particles with the plasma of the lower solar atmosphere. Particle acceleration during reconnection at coronal null points has been explored as a mechanism to explain many observed phenomena, see, e.g., \citep{baumann2013,pallister2021}, and references therein. However, none of these previous studies addressed particle acceleration during null creation/annihilation processes, which will be an important topic for future study.

As magnetic reconnection is ubiquitous in the solar corona and serves as the underlying cause for the generation of 3D nulls, it can explain their abundance in the solar atmosphere. The nulls, being preferential sites of reconnection, a continuous process of null generation and the corresponding reconnection can, in principle, explain their contribution to chromospheric and coronal heating. A concrete validation of this proposal requires an assessment of the energy dynamics involved, but it remains fascinating nonetheless.

Moreover, the fact that nulls are ordered entities, possessing spine and fan field lines, suggests their spontaneous creation as self-organized configurations. The conclusion that magnetic reconnection creates 3D nulls, and that these nulls are self-organized structures, has far-reaching consequences for our understanding of nature and attracts further research.\\

\noindent \textbf{SUPPLEMENTARY MATERIAL}\\
Supplementary materials are provided wherever it is required to support and/or complement the findings.\\

\noindent \textbf{ACKNOWLEDGEMENT}\\
The computations are performed on the Param Vikram-1000 High Performance Computing Cluster of the Physical Research Laboratory (PRL), India. We also wish to acknowledge the visualization software VAPOR (\url{www.vapor.ucar.edu}), for generating the relevant graphics and F. Chiti \emph{et al.} for developing the trilinear null detection technique used here, link of technique is attached (\url{https://zenodo.org/record/4308622#.YByPRS2w0wc}). The corresponding theory can be found in \citet{2007PhPl...14h2107H}.

\noindent \textbf{DATA AVAILABILITY}\\
The simulation data that supports the findings of this study is available from the corresponding author upon reasonable request.


  


\bibliography{ref}

\end{document}